\documentclass[traditabstract]{aa}  
\usepackage{graphicx,natbib, txfonts}
\usepackage[normalem]{ulem}
\usepackage{colortbl}

\def\kms {\hbox{km\,s$^{-1}$}}
\def\ergs {\hbox{erg\,s$^{-1}$}}
\newcommand{\e}[1]{\times 10^{#1}}
\newcommand{\iso}[1]{${}^{#1}$}

\newcommand{\wl}{$\lambda$}
\newcommand{\msun}{M$_\odot$}
\newcommand{\wll}{$\lambda \lambda$}

 \newcommand{\ccm}{{\rm \ cm}^{-3}}

\setcitestyle{round,aysep={},yysep={,}}

\begin{document}

\title{The progenitor mass of the Type IIP supernova SN 2004et from late-time spectral modeling}

   \author{Anders Jerkstrand\inst{1,2} 
          \and
           Claes Fransson\inst{1}
          \and
           Kate Maguire\inst{3}
          \and         
          Stephen Smartt\inst{2}
          \and
          Mattias Ergon\inst{1}
          \and
          Jason Spyromilio\inst{4}
         }


   \institute{Department of Astronomy, The Oskar Klein Centre, Stockholm University, SE-106 01 Stockholm, Sweden \\
            \email{andersj@astro.su.se}       	
    \and Astrophysics Research Centre, School of Maths and Physics, Queen's University Belfast, Belfast BT7 1NN, UK
   \and Department of Physics (Astrophysics), University of Oxford, DWB, Keble Road, Oxford OX1 3RH, UK
   \and ESO, Karl-Schwarzschild-Strasse 2, 85748 Garching, Germany
   }
   \date{Submitted 3 May 2012 / Accepted 8 August 2012.}

  \abstract{
  SN 2004et is one of the nearest and best-observed Type IIP supernovae, with a progenitor detection as well as good photometric and spectroscopic observational coverage well into the nebular phase. Based on nucleosynthesis from stellar evolution/explosion models we apply spectral modeling to analyze its $140-700$ day evolution from ultraviolet to mid-infrared. We find a $M_{\rm ZAMS}= 15$ \msun~progenitor star (with an oxygen mass of 0.8 \msun) to satisfactorily reproduce [\ion{O}{i}] \wll6300, 6364 and other emission lines of carbon, sodium, magnesium, and silicon, while 12 \msun~and 19 \msun~models under- and overproduce most of these lines, respectively. This result is in fair agreement with the mass derived from the progenitor detection, but in disagreement with hydrodynamical modeling of the early-time light curve. 
From modeling of the mid-infrared iron-group emission lines, we determine the density of the ``Ni-bubble'' to $\rho(t) \simeq 7\e{-14}\times(\mbox{t}/100~\mbox{d})^{-3}$ g cm$^{-3}$, corresponding to a filling factor of $f = 0.15$ in the metal core region ($V = 1800$ \kms). We also confirm that silicate dust, CO, and SiO emission are all present in the spectra.}


\maketitle


\section{Introduction}
Stars with Zero Age Main Sequence (ZAMS) mass greater than about 8 \msun~end their lives as core-collapse supernovae (SNe).
Over half of these events (per unit volume) are classified as Type IIP \citep{Li2011}, showing hydrogen lines, as well as a $3-4$ month plateau in the light curve, implying the presence of a massive hydrogen envelope. 
After the plateau phase, the core of the SN becomes visible, glowing from radioactive input by \iso{56}Co. As long as the ejecta remain opaque to the gamma-rays emitted in the decay, the light curve follows the exponential decay of \iso{56}Co, with an $e$-folding time of 111.4 days. The spectrum evolves from being dominated by absorption lines and lines exhibiting P-Cygni profiles superimposed upon a blackbody-like continuum to strong emission lines and a weaker continuum. As the temperature falls, thermal emission eventually shifts into the infrared, but ultraviolet/optical features remain, caused by non-thermal ionizations and excitations.

From hydrodynamical modeling, the progenitors of Type IIP SNe are believed to be red supergiants (RSGs) \citep{Chevalier1976, Falk1977}. 
The rapidly growing image archives have improved the prospects of confirming this by identifying SN progenitors in imaging surveys. Such identifications have now been made for the Type IIP SNe SN 2003gd, SN 2004A, SN 2004et, SN 2005cs, SN 2008bk, SN 2008cn, and SN 2009md \citep[][and references therein]{Smartt2009a, EliasRosa2009, Fraser2011}. For SN 2003gd, the progenitor identification was verified by follow-up observations showing that the candidate had disappeared \citep{Maund2009}. In the cases where multiple-band detections were made, the progenitors have shown properties consistent with the RSG hypothesis, except for SN 2008cn, where the progenitor was yellow\footnote{There is, however, the possibility that the detected source is a blend of two or more stars, and SN 2008cn differs from normal Type IIP SNe in its plateau length.} \citep{EliasRosa2009}.

The estimated luminosity of the progenitor allows its ZAMS mass to be determined from stellar evolution models. The resulting values are consistently found to be below 20 \msun, with a statistical analysis yielding a range of $8.5_{-1.5}^{+1}-16.5_{-1.5}^{+1.5}$ \msun~for the progenitor population \citep{Smartt2009b}, although dust effects may produce a somewhat higher upper boundary \citep{Walmswell2012}. The fate of stars in the $20-25/30$ \msun~range then remains unclear, as standard stellar evolution models predict also these to evolve to RSGs and explode as Type IIP SNe \citep[e.g.,][]{HegerLanger2000, Meynet2003, Eldridge2008}. Models including rotation show that stars with initial mass greater than 20 M$_{\odot}$ and time-averaged equatorial velocities of $\gtrsim$200\,km s$^{-1}$ could lose much of their hydrogen envelope and move bluewards in the HR diagram \citep[e.g.][]{Meynet2003}. While this is immediately attractive as an explanation for the lack of luminous RSGs as Type IIP progenitors, one must account for the broad observed distribution of equatorial velocities \citep[e.g.][]{Hunter2008} as well as the lack of detection of blue, luminous progenitors for Type IIn/Ib/Ic SNe \citep{Smartt2009a}. 

Another method to derive information about the exploded star is through radiation-hydrodynamical modeling of the bolometric light curve. \citet{Litvinova1983, Lit1985} presented scaling relations from fits to a grid of Type IIP SN models allowing for the determination of the explosion energy, ejecta mass, and progenitor radius from the observed plateau length, $V$-band magnitude, and photospheric velocity. However, the neglect of \iso{56}Ni in these models renders them viable only for SNe with very small \iso{56}Ni masses. \citet{Eastman1994} found that a (typical) \iso{56}Ni mass of 0.06 \msun~extends the plateau by 40\%, and \citet{Kasen2009} found a prolongation by 20-30\% for \iso{56}Ni-masses of $0.05-0.1$ \msun. Since the ejecta mass has a strong scaling with the plateau length (to the power 2.9) in the \citet{Lit1985} relations, a significant overestimate of the ejecta mass results. 

More reliable results can be obtained by fitting hydrodynamical models including \iso{56}Ni to the whole light curve evolution. 
Such modeling has been undertaken for SN 1999em \citep{Baklanov2005, Utrobin2007, Bersten2011}, SN 2003Z \citep{UCP2007}, SN 2004et \citep{Utrobin2009}, SN 2005cs \citep{Utrobin2008}, and SN 2009kf \citep{Utrobin2010}, obtaining 
ejecta masses $M_{\rm ej}=14-28$ \msun.
As discussed by \citet{Utrobin2009} and \citet[][M10 hereafter]{Maguire2010}, these ejecta masses are generally too high to be consistent with the initial masses determined from direct observations of SN progenitors, as well as for what is expected from stellar evolution in general. Recently, \citet{Inserra2011} and \citet{Inserra2012}, using the radiation hydrodynamics code of \citet{Pumo2011}, determined lower ejecta masses of $M_{\rm ej}=5-7.5$ and $M_{\rm ej}=8-12$ \msun\ for the Type IIP SNe 2007od and SN 2009bw.

A third method for diagnosing the progenitor mass is through late-time spectral modeling. In the nebular phase ($t \gtrsim 150$ days), the inner ejecta become visible, and the various nuclear burning zones can be analysed. Stellar evolution models predict the metal core mass to strongly increase with progenitor ZAMS mass \citep[e.g.][]{WW95}, making it possible to distinguish between different progenitors by determining the nucleosynthesis yields. Emission lines of carbon, oxygen, neon, sodium, magnesium, silicon, and sulphur are the main signatures that can potentially constrain the progenitor mass.

Despite being the most common core-collapse SN class, there are so far only a handful of Type IIP SNe that have extensive temporal and spectral coverage in the nebular phase, and not much information has as yet been derived from these observations. The best existing datasets are for SN 1990E \citep{Schmidt1993, Gomez1993, Benetti1994}, SN 1999em \citep{Leonard2002,Elmhamdi2003}, SN 1999gi \citep{Leonard2002gi}, SN 2002hh \citep{Pozzo2006}, SN 2004dj \citep{Chugai2005}, SN 2004et \citep[][M10]{Sahu2006}, SN 2005cs \citep[e.g.][]{Pastorello2009}, 2007it \citep{Andrews2011}, and 2007od \citep{Inserra2011}. Of these, only SN 1999em was well monitored also in the near-infrared (NIR). 

Apart from the sparsity of data, spectral analysis needs complex tools to model the transformation of energy from radioactive decay to emergent UVOIR radiation. Temperature and NLTE solutions for a large number of zones and elements, including non-thermal heating/ionization/excitation rates, and multi-line radiative transfer are the fundamental ingredients that must be calculated, together making up millions of constraints whose self-consistent solution must be iteratively sought. In addition, molecules, dust, and time-dependent effects may cause further complications.

So far, SN 1987A is the only Type II SN for which detailed spectral modeling has been undertaken.  Despite being a Type IIpec rather than a Type IIP, it was probably similar to a Type IIP in the nebular phase, since the nucleosynthesis is largely unaffected by the late-time evolution of the envelope. \citet{Xu1991} and \citet{Kozma1992} obtained the solutions for the ionization, excitation, and heating produced by the gamma-rays and positrons, which is the first step in the nebular phase modeling.
The evolution of individual lines were analyzed in a series of papers by \citet{Xu1992} (hydrogen), \citet{Li1992} (oxygen), \citet{Li1993iron} (iron, cobalt, nickel), \citet{Li1993} (calcium), and \citet{Li1995} (helium). The thermal evolution of the envelope was modelled in detail by \citet{deKool1998}, and \citet[][in the following KF98a,b]{Kozma1998I, Kozma1998II} computed the spectra from detailed explosion models to study the evolution of temperature, ionization, and line fluxes in the $200-800$ day range. \citet{Kjaer2010} and \citet[][J11 hereafter]{Jerkstrand2011} analyzed the spectrum in the \iso{44}Ti-powered phase ($t \gtrsim 5$ years), including the effects of multi-line radiative transfer.

Modeling of other objects includes the work by \citet{Dessart2011}, who compared the emergent spectra of Type IIP explosion models to SN 1999em, assuming complete thermalization of the gamma-rays, but performing a detailed radiative transfer calculation. 

In this paper we undertake a detailed analysis of  one of the brightest and best-observed Type IIP SNe to date, SN 2004et.
This SN has been subject to progenitor analysis, hydrodynamical modeling, and some qualitative spectral analysis. \citet{Li2005} identified a progenitor candidate in ground-based pre-explosion images. However, \citet {Crockett2011} showed that this candidate was still visible in 2007 (3 years after explosion), and that the source in the original images was a composite of two or three sources. 
They identified an excess flux in the pre-explosion images contributed by the true SN progenitor, and for two different assumptions about the late-time SN flux determined $M_{\rm ZAMS}=8$ and 10 \msun, with an error of $+5$/$-1$ \msun. With new post-explosion imaging and updated methods to determine bolometric corrections and synthetic colours from model spectra, Fraser et al. 2012 (in prep.) have revised this to $M_{\rm ZAMS}=11_{-1}^{+2}$ \msun. The best estimates for luminosity, temperature, and radius are log $\left(L/L_\odot\right) = 4.8$, $T= 3600$ K, and $R_0 =  650 ~R_\odot$.
 
Through radiation-hydrodynamical modeling, \cite{Utrobin2009} derived a large ejecta mass of $M_{\rm ej}=24.5\pm 1~M_\odot$, the highest value for any Type IIP modeled in detail so far. They also determined a progenitor radius $R_0=1500\pm 140~R_\odot$, explosion energy $E=2.3\pm 0.3~\mbox{B}$ (1 B = $10^{51}$ ergs) and \iso{56}Ni mass $M(^{56}\mbox{Ni})=0.068\pm 0.009 M_\odot$.

Spectroscopic analysis is so far limited to some qualitative results from the nebular spectrum by \citet[][S06 hereafter]{Sahu2006} and M10. 
S06 found the [\ion{O}{i}] \wll6300, 6364 luminosity to be of comparable to the one in SN 1987A, and from the similar \iso{56}Ni mass they conclude that SN 2004et should have a similar oxygen mass. From this follows a similar progenitor mass, which was $16-22$ \msun~ for SN 1987A \citep{Arnett1989}. 
M10 found the [\ion{O}{i}] \wll6300, 6364 flux normalized to the \iso{56}Ni mass to be 15\% lower than in SN 1987A at 285 days, and assuming all other things equal the oxygen mass would then be smaller by the same factor. With the strong dependency of oxygen production on progenitor mass \citep[e.g.][]{WW95}, SN 2004et would then be expected to have a progenitor mass only slightly lower than SN 1987A. 

Taken together with the progenitor analysis and light curve modeling, there are therefore conflicting results for how massive the progenitor of SN 2004et was. The aim in this paper is to shed more light on this issue by modeling the photometric and spectroscopic nebular phase data. To this end, we use the spectral synthesis code described in J11 (see also \citet{Maurer2011} for testing of the code), with some modifications described in Appendix \ref{sec:updates}. We compare the model spectra produced by using three different explosion models as input  (from $M_{\rm ZAMS}=$ 12, 15 and 19 \msun\ progenitors), and investigate which produces best agreement with observations. This paper is complemented by a more general discussion of the nebular phase spectra of Type IIP SNe in \citet[][in the following M12]{Maguire2012}.

\section{Observational summary}
SN 2004et was discovered on Sep 27 2004 \citep{Zwitter2004}, with the explosion date determined to Sep 22 \citep{Li2005}. The host galaxy was the nearby (5.5 Mpc) starburst galaxy NGC 6946, which has hosted nine SNe since 1917. The SN was classified as a Type IIP, showing broad P-Cygni Balmer lines \citep{Zwitter2004,Filippenko2004}.  
Intensive photometric and spectroscopic monitoring was initiated, and is described in S06, \citet[][M07 hereafter]{Misra2007}, \citet[][K09 hereafter]{Kotak2009} and  M10. The luminosity was among the highest of all Type IIP events, both on the 110 day plateau and in the radioactive tail phase (M10). 

The SN was detected at radio wavelengths on October 5 2004 \citep{Stockdale2004}, implying the presence of significant amounts of circumstellar material. Follow-up observations were made with MERLIN \citep{Beswick2004}, GMRT (M07) and VLBI \citep{Marti-Vidal2007}. X-ray and radio emission \citep{Rho2007, Marti-Vidal2007} indicate the presence of a progenitor wind. A mass loss rate of $\dot{M}=10^{-5}$ \msun~yr$^{-1}$ was estimated from the radio turn-on time, assuming a wind velocity of $v_{\rm w}=10$ \kms~and a circumstellar temperature of $T_{\rm cs}=10^5$ K~\citep{Chevalier2006}\footnote{The inferred mass-loss rate scales as $\dot{M}\propto v_{\rm w} T_{\rm cs}^{3/4}$.}, whereas \citet{Rho2007} found $\dot{M} = 2\e{-6}$ \msun~yr$^{-1}$ from modeling of the X-ray emission.

K09 obtained Spitzer mid-infrared (MIR) spectra at $5-35$ $\mu$m for seven epochs between 295 and 1385 days, as well as optical Keck spectra at days 823, 933 and 1146. The late optical spectra appear to be dominated by circumstellar interaction in the form of wide ($\mbox{HWZI}= 8500~\mbox{\kms}$), box-shaped H$\alpha$, [\ion{Ca}{ii}] \wl7300, and [\ion{O}{i}] \wll 6300, 6364, as well as narrow (70 \kms) H$\alpha$ and [\ion{N}{ii}] \wll 6548, 6583. Spectroscopic identification of silicate dust is reported, with a mass of a few times $10^{-4}$ \msun. Further MIR data and analysis is presented by \citet[][F11 hereafter]{Fabbri2011}.

\subsection{Distance and extinction}
\label{sec:ext}
We adopt a distance of 5.5 Mpc for NGC 6946, as determined using the \ion{H}{i} Tully-Fisher relation \citep{Pierce1994}, and in agreement with the average value of several other distance determinations \citep{Schmidth1993, Schoniger1994, Karachentsev2000, Sahu2006}. 

We adopt an extinction of $E_{\rm B-V}=0.41$, based on the sum of local and foreground \ion{Na}{i} D absorption lines \citep{Zwitter2004}. 
The galactic foreground extinction towards NGC 6946 is $E_{\rm B-V}=0.34 $ \citep{Schlegel1998}. 

\subsection{Metallicity}
\label{sec:metallicity}
Metallicity measurements would traditionally come from the [\ion{O}{ii}] and [\ion{O}{iii}] line fluxes in nebular or \ion{H}{ii} regions at, or close to the SN position \citep[e.g.,][]{Smartt2009b, Modjaz2011}. From the radial position of SN 2004et ($r_{\rm g} = 0.92r_{\rm 25}$), and the gradient of \citet{Pilyugin2004}, \citet{Smartt2009b} estimated the oxygen abundance of the progenitor to $12+\log{\rm O/H}=8.3$\,dex, which is a factor of about two lower than solar metallicity. However the determination of absolute oxygen abundances from the strong line ratio of [\ion{O}{ii}] and [\ion{O}{iii}] has
been plagued by large differences between the calibrations employed. As \citet{Moustakas2010} show, the oxygen abundance at the position
of SN 2004et determined by two physically motivated calibrations is 8.9 and 8.3\,dex. This large
difference of a factor four makes it difficult for us to choose a specific stellar metallicity for the progenitor model. The solar value
of 8.7\,dex \citep{Asplund2009} is intermediate between the two and as we have no clear way of distinguishing between the discrepant
results we employ a solar metallicity model. We discuss the influence of metallicity on the modeling in the discussion (Sect. \ref{sec_disc}).


\subsection{Photometry}
We use $UBVRI$ photometry presented by S06, M07, M10 and F11. S06 obtained photometry for $8-541$ days, M07 for $14-90$ and $370-470$ days,
and M10 for $7-1214$ days. F11 obtained $VRI$ photometry for $79-1412$ days. 

For nebular-phase NIR photometry, we use M10 measurements of $J$ and $H$ magnitudes on days 134 and 141, and $JHK$ magnitudes from F11 for days 268, 317, 390, and 652. We also use day 307 $JHK$ observations by M10, which due to their importance for the only NIR spectrum taken were recalibrated with an extended set of reference stars. The raw data were reduced following standard
procedures with the addition of corrections for quadrant crosstalk and
field distortion. We determined colour constants from standard fields and approximate
S-corrections from the standard star spectrum and the filter transmission curves, to as accurately as possible transform from the instrumental system to the standard (2MASS) system. We find $J=15.70\pm 0.03$, $H=15.48\pm 0.02$, and $K=15.12\pm0.03$.

In the MIR, we use 3.6 $\mu$m, 4.5 $\mu$m, and 24 $\mu$m Spitzer photometry from K09.

\subsection{Spectroscopy}
\label{sec:dataspec}

We use a set of sixteen optical, one near-infrared, and three mid-infrared spectra from S06, K09, and M10, a summary of which can be seen in Table \ref{table:spectro}. We recalibrated the NIR TNG/Amici spectrum (day 307) using the new photometry reductions described above. The prism used in these observations has very low resolution 
preventing wavelength calibration using lamp or sky lines because of
line blending. Instead the observatory provides a pixel to wavelength mapping to be
calibrated using known features in the spectrum by applying a constant pixel shift.
We used this method to calibrate the spectrum by alignment of the
strongest lines to the model spectra. The raw data were reduced following standard procedures with the addition of corrections for quadrant cross talk and the spectrum then extracted and corrected for telluric absorption using the standard star spectrum. The absolute flux calibration was performed by a least squares fit of synthetic and measured $J$ and $H$ magnitudes (spectral response in the $K$-band region involves significantly higher uncertainties).

All spectra displayed in the paper have been corrected for the host galaxy recession velocity of 48 \kms \citep{Sandage1981}. The optical and NIR spectra have been corrected for extinction with $E_{\rm B-V}=0.41$ (Sect. \ref{sec:ext}) and the \citet{Cardelli1989} extinction law. No correction was applied to the MIR data, for which the extinction is negligible.  



\begin{table}[htb]
\centering
\caption{Summary of spectroscopic observations used in this paper.}
\begin{tabular}{ c c c c c }
\hline
\hline
Phase & Date & Wl. coverage  & Res. & Ref. \\
\hline
163 d  & 2005-03-03 & $3500-9200$ \AA & 7 \AA & S06 \\
169  & 2005-03-09 & $3800-7730$ \AA & 24 \AA & M10 \\
212  & 2005-04-21 & $3500-9200$ \AA & 7 \AA & S06\\
249  & 2005-05-28 & $3500-9200$ \AA & 7 \AA & S06\\
257  & 2005-06-05 & $3800-8130$ \AA & 13 \AA & M10 \\
284  & 2005-07-02 & $4000-7830$ \AA & 36 \AA & M10\\
295  & 2005-07-13 & $5.2-31$ $\mu$m & $\sim 0.1 \cdot \lambda_{\rm \mu m}$ $\mu$m & K09\\
301  & 2005-07-19 & $3500-9200$ \AA & 7 \AA & S06 \\
307  & 2005-07-25 & $0.8-2.54$ $\mu$m &$0.2\cdot \lambda_{\rm \mu m}$ $\mu$m & M10\\
314  & 2005-08-01 & $3500-9200$ \AA & 7 \AA & S06\\
341  & 2005-08-28 & $3400-8000$ \AA & 13 \AA & M10\\
350 & 2005-09-06  & $5.2-35$ $\mu$m &  $\sim 0.1 \cdot \lambda_{\rm \mu m}$ $\mu$m & K09\\
384  & 2005-10-10 & $3500-7730$ \AA  & 24 \AA & M10\\
391  & 2005-10-17 & $3500-9200$ \AA & 7 \AA & S06\\
401  & 2005-10-27 & $3800-7730$ \AA & 25 \AA & M10\\
401  & 2005-10-27& $3500-9200$ \AA  & 7 \AA & S06\\
408  & 2005-11-03 & $4000-7730$ \AA  & 25 \AA & M10\\
428  & 2005-11-23 & $3500-9200$ \AA & 7 \AA & S06\\
451 & 2005-12-16  & $5.2-20$ $\mu$m &  $\sim 0.1 \cdot \lambda_{\rm \mu m}$ $\mu$m & K09\\
465  & 2005-12-30 & $3500-9200$ \AA  & 7 \AA & S06\\
\hline
\end{tabular}
\label{table:spectro}
\end{table}

\section{Modeling}
\label{sec:modeling}
The spectral synthesis code we use is described in detail in J11. 
In summary, we compute the energy deposition by radioactive isotopes by performing an effective opacity transfer of the gamma-rays, and assuming on-the-spot deposition for the leptons and X-rays. The relative fractions of the energy going into heating, ionization, and excitation are calculated as described in \citet{Kozma1992}, and the equations of statistical and thermal equilibrium (including non-thermal heating/ionization/excitation, collisional (thermal) excitation/de-excitation, radiative excitation/de-excitation,  photoionization, radiative recombination, charge transfer, photoelectric heating, free-free heating, line cooling, recombination cooling, and free-free cooling) are then iterated with a Monte Carlo simulation of the radiation field until convergence has been achieved. The radiative transfer treatment uses the Sobolev approximation. The method is a quasi-$\Lambda$-iteration, but differs in that it treats continuum and part of the line scattering on-the-fly in each radiation field simulation, rather than including them in the emissivity functions. This improves the convergence properties by allowing photons to travel longer distances than one mean-free-path in each iteration. The emergent spectrum is obtained by binning the escaping Monte Carlo packets, and smoothing with a Gaussian of $\mbox{FWHM}=600$ \kms.

The code has been updated in some respects compared to the version used in J11. Because the computations here are for earlier epochs than in J11, photoexcitation/de-excitation rates in the NLTE solutions cannot be approximated to zero. We therefore implemented a modified Monte Carlo operator to allow for partial deposition of photo-excitation energy along the photon trajectories (we keep the coupling partial to avoid approaching a classic $\Lambda$-iteration, and keep the NLTE solutions limited in size). The details of the new treatment can be found in Appendix \ref{sec:updates}. We have also added electron scattering, as well as the influence of line overlap on Ly$\alpha$ and Ly$\beta$ Sobolev escape probabilities, also detailed in Appendix \ref{sec:updates}.

\subsection{Explosion model and zoning}
As input we use the 12 \msun, 15 \msun, and 19 \msun~explosion models computed by \citet[WH07 hereafter]{Woosley2007}, which are non-rotating stars of solar metallicity evolved with the KEPLER hydrodynamics code 
and then exploded with a piston, endowing a final kinetic energy of 1.2 B to the ejecta. The mass-loss prescription in these models is based on the empirically established relations by \citet{Nieu1990}, both for the main sequence and the RSG phase. The mass loss for the three stars are 1.2, 2.3, and 3.8 \msun.

We divide the ejecta into eight distinct zone types : Fe/He, Si/S, O/Si/S, O/Ne/Mg, O/C, He/C, He/N, and H, named after their dominant constituents. The mass of the Fe/He zone is set to give a total \iso{56}Ni-mass of 0.062 \msun~in the ejecta (see Sec. \ref{sec:56Nimass}). Table \ref{table:zonemasses} shows the zone masses in the three models we use, and Tables \ref{table:chemcomp1}-\ref{table:chemcomp3} in the appendix show their detailed composition. 

\begin{table}[h!]
\centering
\caption{Zone masses (in $M_\odot$) of the three different explosion models used.}
\begin{tabular}{l c c c}
\hline
\hline
Zone / Model & 12  & 15  & 19\\
\hline
Fe/He   & 0.10 & 0.087 & 0.076 \\
Si/S    & 0.061 & 0.080 & 0.16\\
O/Si/S  & 0.13  & 0.24 & 0.17\\
O/Ne/Mg & 0.14  & 0.45 & 1.9 \\
O/C     & 0.16  & 0.43 & 0.60 \\
He/C    & 0.16  & 0.74 & 0.88 \\
He/N    & 0.92  & 0.43 & 0.40  \\
H       & 7.7   & 8.4  &  9.5\\
\hline
Total & 9.3 & 10.9 & 13.6\\
\hline
\end{tabular}

\label{table:zonemasses}
\end{table}

The explosion models are 1D, and therefore lack the macroscopic mixing which is known to occur \citep[e.g.,][]{Kifonidis2006, Hammer2010}. We therefore distribute the zones over a core, where each zone is distributed as $N_{\rm cl}$ spherical clumps, completely macroscopically mixed in velocity, and an envelope, where the zones are concentric shells, with no mixing. The Fe/He, Si/S, O/Si/S, O/Ne/Mg, O/C, and He/C zones exist only in the core, whereas the He/C, He/N, and H zones exist partially in the core (fractions $x_{\rm He}$ for the He zones and $x_{\rm H}$ for the H zone), and partially in the envelope (fractions $1-x_{\rm He}$ and $1-x_{\rm H}$). The first two shells in the envelope are the unmixed He/C and He/N zones. External to these zones are logarithmically spaced ($V_{\rm i}/V_{\rm i-1} = 1.2$) shells of the H-zone material. The velocities of the two unmixed He-shells are set so their densities equal that of the first hydrogen shell.

The density of the H-envelope is set to follow the density profile of the input model, but scaled down with a constant to account for the mixing of material into into the core. This downscaling was about 10\% for all three models. The envelope is terminated at $10^4$ \kms, as the density decreases steeply beyond a few thousand \kms, and little line or continuum opacity remains at higher velocities than this in the nebular phase.
 
\begin{figure}[htb]
\centering
\includegraphics[trim=5mm 5mm 10mm 12mm, clip, width=0.95\linewidth]{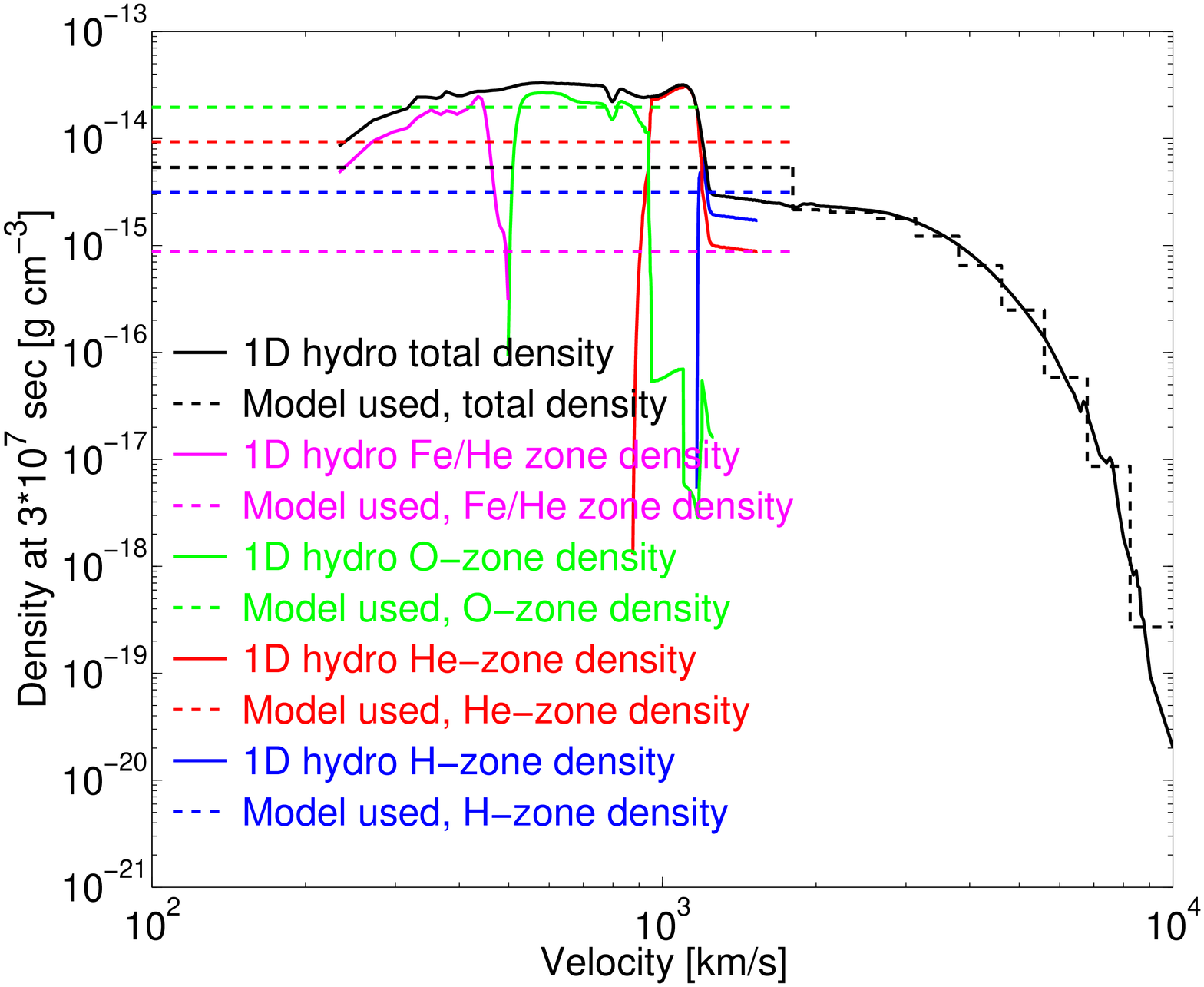}
\caption{The density profile in the 1D hydrodynamical model (solid lines) and in the artificially mixed model used in this paper (dashed lines). Note that in the artificially mixed model, the zones in the core are distributed as clumps with individual filling factors, the densities plotted are the densities within each clump.}
\label{fig:ejectastruct}
\end{figure}

Although our artificial mixing alters the velocity field for each species compared to the 1D input models, the explosion energy of 1.2 B is preserved to within 3\% in all models. Fig. \ref{fig:ejectastruct} shows the velocity field for the 15 \msun\ model that we use, compared to the 1D hydro simulation.

\subsection{Model parameters}
 \label{sec_mod_par}
The parameters of the model are the velocity dividing the core and the envelope ($V_{\rm core}$), the fractions of the He and H zones mixed into the core ($x_{\rm He}$ and $x_{\rm H}$), the filling factors of the different zones in the core ($f_{\rm Fe/He}$, $f_{\rm Si/S}$, $f_{\rm O/Si/S}$, $f_{\rm O/Ne/Mg}$, $f_{\rm O/C}$, $f_{\rm He/C}$, $f_{\rm He/N}$, $f_{\rm H}$), the number of clumps per zone in the core ($N_{\rm cl}$), and the dust optical depth function ($\tau_{\rm d}(t))$, which we assume to be gray. Most of these parameters can be constrained from observations or theoretical arguments, as described below.

\subsubsection{Core velocity $V_{\rm core}$}
The core velocity is to first order found by inspecting the line widths of [\ion{O}{i}] $\lambda \lambda$6300, 6364 and [\ion{Fe}{ii}] \wl7155, which were mainly emitted from the core in SN 1987A (KF98b). The wings of these lines may, however, be produced by primordial oxygen and iron in the envelope (M12), and are therefore unsuitable markers of the core width. A more relevant starting point is the half-width-at-half-maximum, $\mbox{HWHM}$, which is 
$\sim$1200 \kms~for the [\ion{O}{i}] \wll 6300, 6364 lines (M12). Inspection of the [\ion{Fe}{ii}] \wl7155 line gives similar values. The emission line profile from a homogeneous sphere expanding homologously is parabolic with $V_{\rm HWHM}=0.71 V_{\rm core}$, so $V_{\rm core}$ can be estimated to 1700 \kms. We find that a value $V_{\rm core}=1800~\kms$ best reproduces the line profiles, and is chosen for the modeling.

\subsubsection{In-mixing fractions $x_{\rm He}$ and $x_{\rm H}$.} 
Several 2D and 3D hydrodynamical calculations have shown that $1-2$ \msun~of the H-envelope gets mixed with core material in the explosion  \citep{Herant1991,Mueller1991,Herant1994,Kifonidis2006,Hammer2010}. This was also found from modeling of the H$\alpha$ evolution in SN 1987A (KF98b). \citet{Herant1994} obtain 1.4 \msun~of H (20\% of the total) and 1.3 \msun~of He (85\% of the total) inside 1800 \kms, for the 2D explosion of a 12 \msun~RSG. The 3D explosion simulation of a 15.5 \msun~blue supergiant by \citet{Hammer2010} shows 60\% of the He and and 15\% of the H mixed inside $V_{\rm core}=1800$ \kms. Based on this we adopt  $x_{\rm He}=0.60$ and $x_{\rm H}=0.15$.

\subsubsection{Filling factors}
The filling factor of the iron zone is determined by the energy input from the \iso{56}Ni decay during the first weeks, creating an ``iron bubble'' in the core \citep{Woosley1988, Herant1991,Basko1994}. As we show in Sect. \ref{sec:izff}, analysis of the Fe/Ni/Co fine-structure lines can be used to determine the filling factor of this zone. We find $f_{\rm Fe/He}\approx 0.15$, which is used throughout the rest of the paper. 

The oxygen density can be derived from the evolution of the [\ion{O}{i}] \wl6300/6364 line ratio, which is found to be similar to the one in SN 1987A (M12). We therefore use the same value for the oxygen density as derived for SN 1987A,  $n_{\rm O}=6.2\e{10}~\mbox{cm}^{-3}$~at 100 days \citep{Spyromilio1991, Li1992}. This determines $f_{\rm O/Si/S}$, $f_{\rm O/Ne/Mg}$, and $f_{\rm O/C}$, giving a scaling proportional to the oxygen mass.

The Si/S zone was found by J11 to have expanded to a low density in SN 1987A, and is given a density ten times lower than the oxygen zone density.

 The remaining core volume is divided between the mixed-in He/C, He/N, and H components to give them equal number densities. 

The resulting filling factors for the three models are listed in Table \ref{table:ff}.

\begin{table}[htb]
\centering
\caption{Filling factors used for the core zones the three models.}
\begin{tabular}{l c c c}
\hline
\hline
Filling factor \textbackslash\ Model  & 12 \msun & 15 \msun & 19 \msun \\
\hline
$f_{\rm Fe/He}$  & 0.15        & 0.15        & 0.15 \\
$f_{\rm Si/S}$   & $1.6\e{-2}$ & $2.2\e{-2}$ & $4.4\e{-2}$ \\
$f_{\rm O/Si/S}$ & $7.9\e{-3}$ & $1.5\e{-2}$ & $5.8\e{-3}$\\
$f_{\rm O/Ne/Mg}$ & $6.8\e{-3}$ & $2.3\e{-2}$ & 0.10  \\
$f_{\rm O/C}$    & $8.7\e{-3}$ & $2.4\e{-2}$ & $2.8\e{-2}$\\
$f_{\rm He/C}$   & $1.4\e{-2}$ & $7.3\e{-2}$      & $7.2\e{-2}$ \\
$f_{\rm He/N}$  & $0.10$         & $4.4\e{-2}$ & $3.4\e{-2}$\\
$f_{\rm H}$    & 0.69         & 0.65         & 0.57 \\
\hline
\end{tabular}
\label{table:ff}
\end{table}

\subsubsection{Number of clumps $N_{\rm cl}$} Several arguments suggest that the number of clumps produced in SN explosions is high. For SN 1987A, 
\citet{Chugai1994} could reproduce the statistical fluctuations in the [\ion{O}{i}] \wll6300, 6364 lines with 2000 clumps. Multi-dimensional models indicate strong mixing and a high degree of fragmentation \citep{Kifonidis2006,Hammer2010}. The strong \ion{O}{i} \wl1.129 $\mu$m line observed in SN 1987A requires synthesized\footnote{We use the word ``synthesized'' for material made during either hydrostatic or explosive burning. Oxygen is made mainly by hydrostatic burning.} oxygen to be mixed with hydrogen down to very small scales in order for the Bowen fluorescence to occur before Ly$\beta$ is converted to $\mbox{Pa}\alpha$ \citep{Oliva1993}, also supporting a large number of clumps. Based on these results, we choose the number of clumps $N_{\rm cl}=10^3$ for our modeling here. 

\subsubsection{Dust}


Both S06 and M10 note that the [\ion{O}{i}] \wl6300 and H$\alpha$ lines show shifts to the blue at around 300 days, and conclude that dust formed at this time. 
We have extended this analysis of the dust formation process by measuring the position of the line peaks for \ion{Na}{i} \wl5896\footnote{We find the \ion{Na}{i} \wll5890, 5896 emission to be much stronger than \ion{He}{i} \wl5876 in all our models. The \ion{Na}{i} lines are optically thick throughout most of the ejecta for the epochs studied here, so \ion{Na}{i} \wl5890 emission will scatter in \ion{Na}{i} \wl5896, making 5896 \AA~the suitable wavelength choice.}, [\ion{O}{i}] \wl6300, H$\alpha$ \wl6563, [\ion{Fe}{ii}] \wl7155 and [\ion{Ca}{ii}] \wl7303\footnote{This is the peak wavelength if the lines are optically thin (as expected for the primordial calcium dominating the emission (M12)) and have Gaussian shapes with $V_{\rm FWHM}=1300$ \kms.}, for all epochs in the 163--465 day range. 

\begin{figure}[htb]
\centering
\includegraphics[trim=0mm 0mm 10mm 0mm, clip, width=1\linewidth]{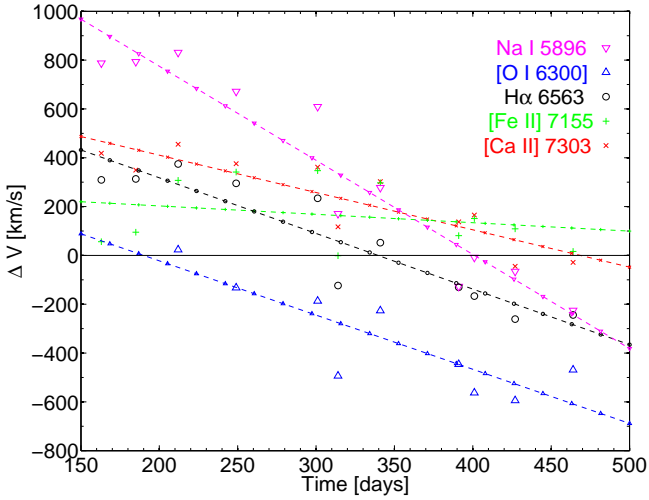}
\caption{Velocity shifts of the peaks of the strongest emission lines in the observed spectra as function of time. The symbols are estimates of the peak positions from fitting a parabola to the part of the line with a flux higher than 85\% of the peak value (95\% for [\ion{Fe}{ii}] \wl7155). The lines are linear fits to the measured values.}
\label{fig:lineshifts}
\end{figure}

We measured the positions of the peaks in two ways; first by determining the wavelength of the maximum (within $\pm$ 2000 \kms~of the rest wavelength), and secondly by fitting a second-order polynomial to the part of the line that is within 15\% of the peak flux. The two methods gave similar results. The resulting measurements with the second method are plotted in Fig. \ref{fig:lineshifts}, together with straight-line fits. 

All emission lines show a continuous blue-shifting trend with time. However, all lines, except [\ion{O}{i}] \wl6300, 
show a \emph{redshift} of a few hundred \kms~at early epochs. One possibility is that the SN has a higher recession velocity than the galactic one of 45 \kms. However, positioned far out from the center in a galaxy seen almost face-on, a peculiar velocity of several hundred kilometers per second is unlikely. Another possibility is that these redshifts are due to electron scattering in the ejecta, whose optical depth is of order unity in the early nebular phase \citep{Chugai1977, Dessart2011}. However, while electron scattering will produce a distinct red wing in the line profile \citep{Auer1972, Chugai1977, Fransson1989}, \citet{Fransson1989} did not find any redshift of the peak in their simulations. We therefore favor a third possibility, that a fundamental asymmetry of the ejecta is the explanation. 

Whatever the reason is for the early red-shifts, later on the line peaks change into blue-shifts, which is most likely caused by dust. The average value of the slope of the linear fits is -2.0 \kms/~day.
Denoting the velocity shift as function of time by $\Delta V(t)$, and taking the dust formation to begin at day 250, we then write
\begin{equation}
\Delta V(t) = -2.0\times (t-250~\mbox{d})~\mbox{km}~\mbox{s}^{-1},~~t > 250~\mbox{d}~.
\end{equation}

\begin{figure}[htb]
\centering
\includegraphics[trim=0mm 0mm 10mm 0mm, clip, width=1\linewidth]{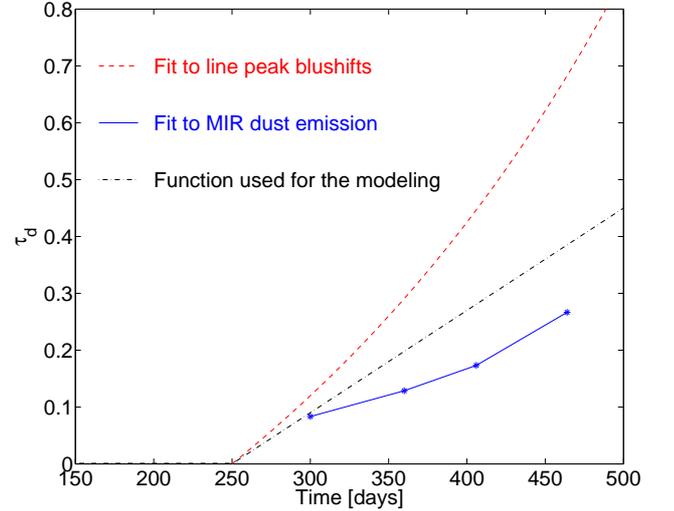}
\caption{The (gray) dust optical depth function $\tau_{\rm d}(t)$ derived from the observed dust luminosity (blue, solid), from the blue-shifts of line profiles (red, dashed), and the one we use for our modeling here (black, dot-dashed).}
\label{fig:dusttau}
\end{figure}

The dust optical depth $\tau_{\rm d}(t)$ can then be estimated from the formula for a uniformly emitting and absorbing sphere of velocity $V_{\rm d}$ \citep{Lucy1991part2}:
\begin{equation}
\frac{\Delta V(t)}{V_{\rm d}} = -1 + \frac{\ln\left[{1+\tau_{\rm d}(t)}\right]}{\tau_{\rm d}(t)}~.
\label{formula:taud}
\end{equation}
The solution for $\tau_{\rm d}(t)$, for $V_{\rm d}=1800$ \kms, is plotted in Fig. \ref{fig:dusttau} (red dashed line). 

The dust hypothesis is supported by Spitzer observations of a warm MIR-continuum at 500 K (K09, F11). In K09, a triple-component black-body fit reveals a steadily increasing fraction of the \iso{56}Co-deposition being re-radiated by a warm component, which is attributed to dust formed in the ejecta. By equating
\begin{equation}
\frac{L_{\rm warm}(t)}{L_{\rm dep}(t)} = 1 - p(t)
\end{equation}
where $L_{\rm warm}(t)$ is the luminosity of the warm component, $L_{\rm dep}(t)$ is the energy deposition, and $p(t)$ is the escape probability from a uniformly emitting sphere with constant opacity \citep{OB},
\begin{equation}
p(t) = \frac{3}{4\tau_{\rm d}(t)}\left[1 - \frac{1}{2\tau_{\rm d}(t)^2}+\left(\frac{1}{\tau_{\rm d}(t)}+\frac{1}{2\tau_{\rm d}(t)^2}\right)e^{-2\tau_{\rm d}(t)}\right]~,
\label{eq:escapeprob}
\end{equation}
the dust optical depth function $\tau_{\rm d}(t)$ can be determined. The resulting values (with $L_{\rm warm}(t)$ and $L_{\rm dep}(t)$ taken from K09) are plotted in Fig. \ref{fig:dusttau} (blue, solid line) \footnote{We only include the results up to day 465, since after that time dust emission from circumstellar interaction affected the MIR emission (K09).}, and are compared with the values derived from the line-shifts. Within the errors expected from the uniform sphere assumption, the independent derivations agree reasonably well. 

For our modeling, we employ a functional form for the dust optical depth lying between the Spitzer and line-shift fits, given by
\begin{equation}
\tau_{\rm d}(t) = 1.8\e{-3}\left(t-250~\mbox{d}\right),~~t>250~\mbox{days}~,
\end{equation}
shown as the black dot-dashed line in Fig. \ref{fig:dusttau}. The extrapolation after 500 days is uncertain, but is not relevant for any of the conclusions in this paper.

\subsection{\iso{56}Ni mass}
\label{sec:56Nimass}
The \iso{56}Ni mass can be determined by comparing the estimated bolometric luminosity in the early tail-phase with the theoretical value of fully trapped \iso{56}Co deposition, which is given by
\begin{equation}
L_{^{56}\rm Co}(t) = 9.92\e{41}\frac{M_{^{56}\rm Ni}}{0.07~M_{\odot}}\left(e^{-t/111.4~d}-e^{-t/8.8~d}\right)~\mbox{erg s}^{-1}~.
\label{eq:56co}
\end{equation}
 Apart from the assumption of full trapping, one also has to assume that the deposited energy is instantaneously re-emitted, and that no other energy source has any influence. 

 We used the $UBVRIH$ photometry from day 142 (M10), together with an extrapolation outside the $H$-band (see below) to estimate the bolometric luminosity at this epoch. 
The linearly interpolated, dereddened $UBVRIH$ flux at 142 days is $1.82\e{41}$ \ergs. The $UBVRIJHKLMN_1N_2N_3Q_0$ photometry of SN 1987A at day 135 \citep{Hamuy1988, Bouchet1989} shows 26\% of the total (linearly interpolated) $U-Q_0$ flux to be outside the $U-H$ range. Assuming a fraction of 0.26 also for SN 2004et, the extrapolated bolometric $U-Q_0$ flux is $2.46\e{41}$ \ergs. From Eq. (\ref{eq:56co}), this corresponds to a \iso{56}Ni mass of 0.062 \msun. This is in good agreement with values derived by S06, M07, and M10.

\subsection{Molecules}

There are now a handful of detections of CO in core-collapse SNe, suggesting that its formation is common \citep[e.g.][]{Gerardy2000, Gerardy2002}. 
In SN 1987A, CO emission was detected already in the photospheric phase, 
and persisted to over 600 days \citep{Spyromilio1988, Bouchet1993}. 

Also SiO was formed in SN 1987A, detected from day $\sim$250 and persisting to over 600 days \citep{Roche1991, Bouchet1993}. It has fewer detections in other SNe compared to CO, likely because it lacks a strong spectral feature in the NIR. 

The 2.3 $\mu$m first overtone band of CO was detected in SN 2004et in the NIR spectrum taken at 307 days (M10). No earlier nebular NIR spectra exist to determine exactly when this feature emerged. 

The 8 $\mu$m fundamental band of SiO was detected in the first Spitzer spectrum at day 295 (K09). It could then be followed until at least day 450, after which it became difficult to distinguish. No spectra covering the fundamental band of CO ($4.6$ $\mu$m) were taken, but its presence is revealed by the photometry (see Sect. \ref{sec:mir}).

Models show that CO forms in the O/C zone, and SiO forms in the O/Si/S zone, with little formation of either in the O/Ne/Mg zone \citep{Liu1995}. Our models do not include molecules, and because these likely determine the temperature of the O/C and O/Si/S zones \citep{Liu1995}, we need some prescription for how to set these temperatures. 

\citet{Spyromilio1988} found, assuming LTE and optically thin conditions, the temperature of the CO clumps in SN 1987A to decline from 3000 K to 1200 K between 192 and 350 days. \citet{Liu1992} found that inclusion of optical depth and NLTE effects gave a temperature evolution of $4000-1900$ K. We assume that the temperature in the CO zone follows a linear interpolation of this result, but levels off at 1000 K at late times
\begin{equation}
T_{\rm O/C-zone}(t) = \mbox{max}\{4700 - 13\cdot\left(t-140~\mbox{d}\right)\mbox{K}, 1000~\mbox{K}\}~.
\end{equation}
For SiO, \citet{Liu1994} found a temperature evolution from 2000 K to 1000 K between 257 and 519 days in SN 1987A. We use the same interpolation method here
\begin{equation}
T_{\rm O/Si/S-zone}(t) = \mbox{max}\{2400 - 3.8\cdot\left(t-140~\mbox{d}\right)\mbox{K}, 1000~\mbox{K}\}~.
\end{equation}
Our results are not sensitive to the exact prescription for the temperature of these zones as they have too small mass and are too cold (assuming that some level of molecular cooling occurs) for any significant atomic thermal emission. 

\section{Results}

\subsection{Temperature evolution}

\begin{figure}[htb]
\centering
\includegraphics[trim=0mm 0mm 10mm 0mm, clip, width=1\linewidth]{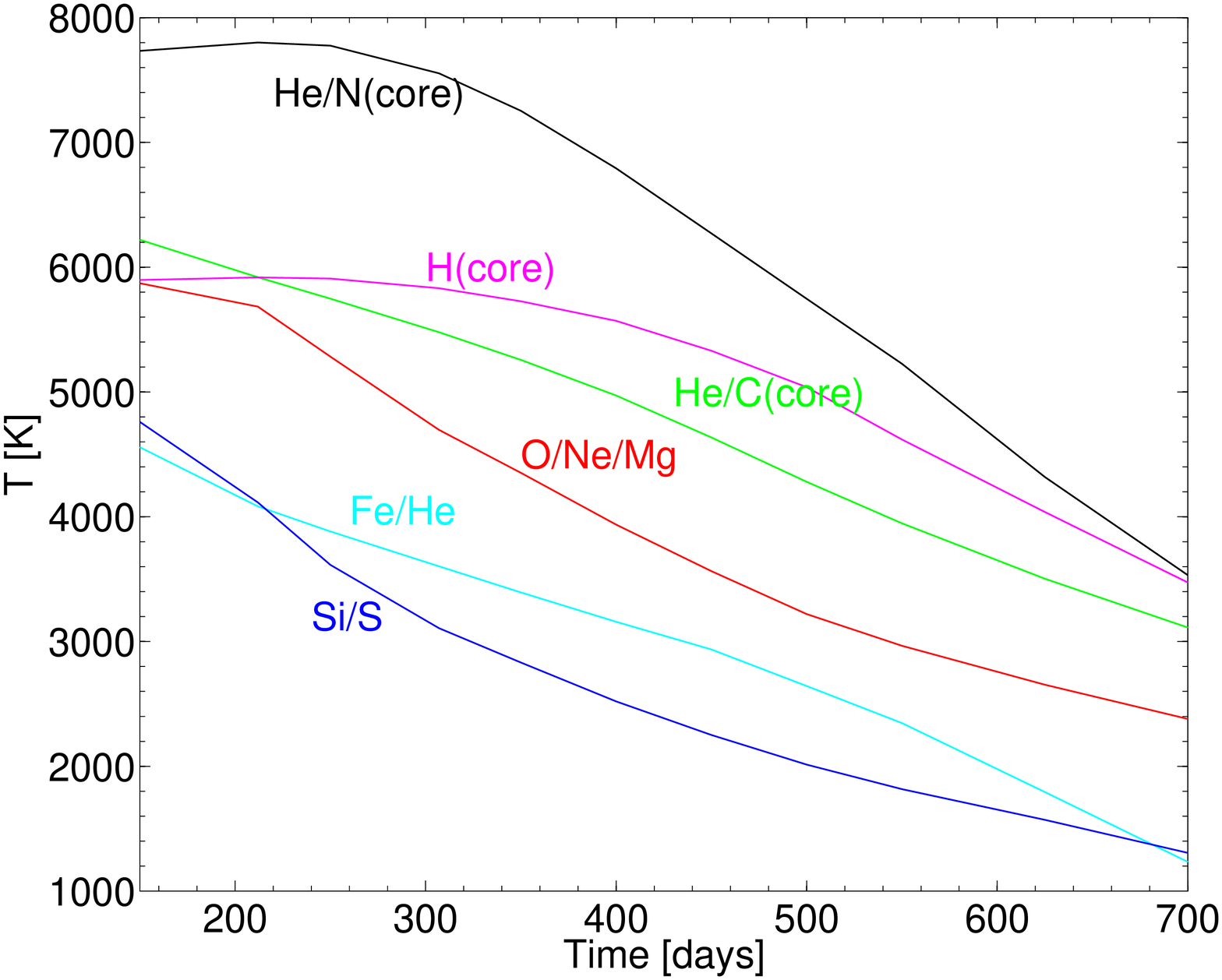}
\caption{Evolution of model temperatures in the various core zones in the 15 \msun\ model.}
\label{fig:temps}
\end{figure}

For later guidance in the interpretation of line fluxes, Fig. \ref{fig:temps} shows the model temperature evolution for the different zones, for the 15 \msun~model. These results may also be of use in interpreting thermal emission line luminosities in other SNe with similar \iso{56}Ni mass.

\begin{figure*}[htb]
\centering
\includegraphics[trim=15mm 10mm 2mm 0mm, clip, width=1\linewidth,height=12cm]{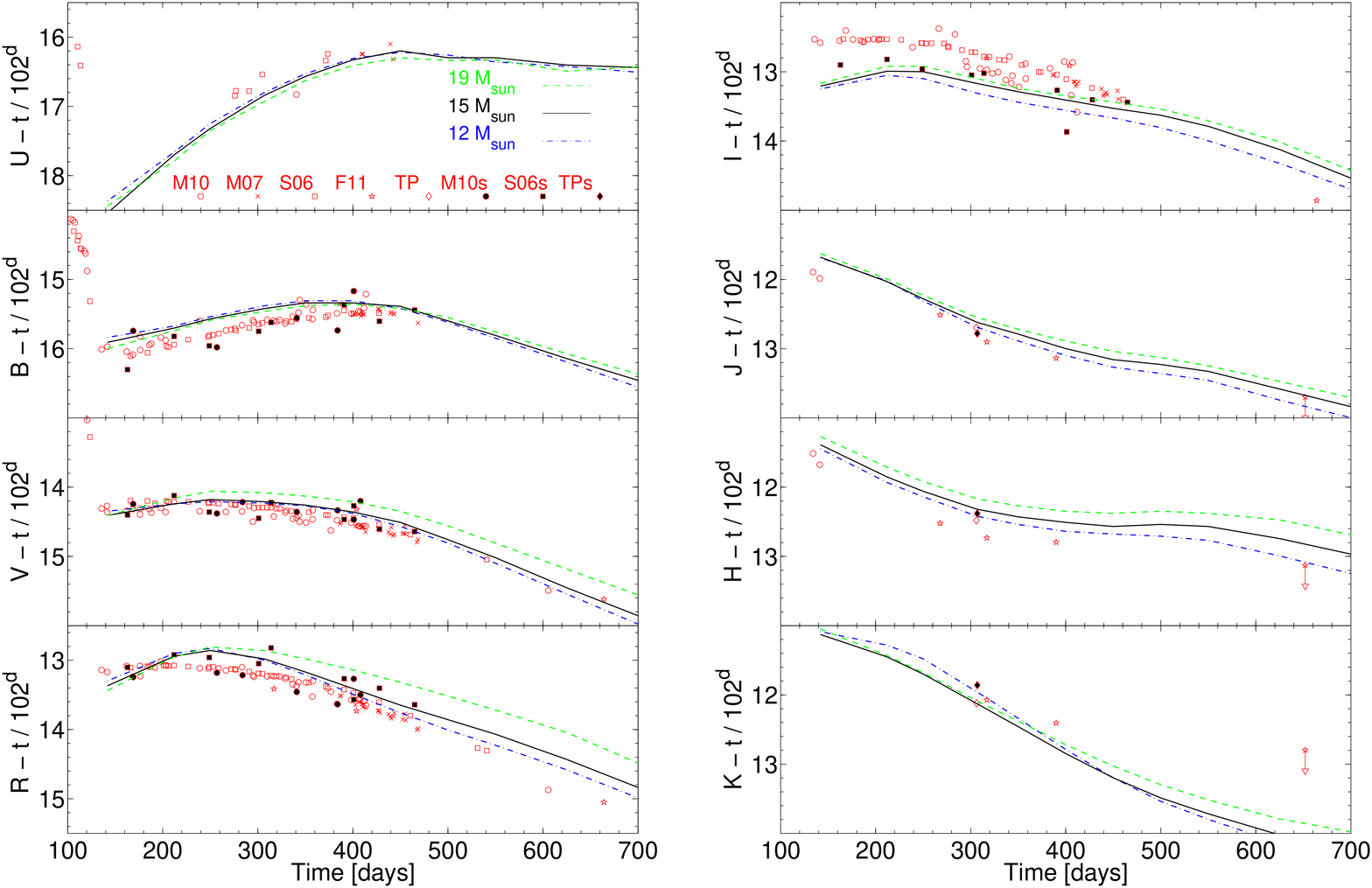}  
\caption{Normalized photometric evolution (all magnitudes have been subtracted by $t/102$ d  to normalize to the \iso{56}Co decay) of the (reddened) 12 \msun~(blue, dot-dashed), 15 \msun~(black, solid), and 19 \msun~(green, dashed) models, compared to observations (red points). T.P. means 'this paper'. The M10s, S06s, and TPs data points are synthetic photometry from the observed spectra.}
\label{fig:photometry}
\end{figure*}

\subsection{Photometric evolution}
\label{sec:phot}

Fig. \ref{fig:photometry} shows the observed (normalized) $U, B, V, R, I, J, H$ and $K$ photometric evolution compared to the models. For clarity 
we have divided all fluxes with the exponential decay factor for \iso{56}Co, 
which in magnitudes corresponds to a subtraction by $t/98~\mbox{d}$. This reduces the span of magnitudes from $\sim$8 to $\sim$2, allowing greater detail to be seen. 

In addition to the observed photometry, we also plot photometry obtained by convolving the observed spectra with the standard filter functions (filled (black) symbols). As is clear from the $R$-band and $I$-band plots, significant differences may occur here, due to the difficulty of transforming from telescope filters to standard filters for sources with strong emission lines, such as nebular phase SNe.

The models differ little in the $U$-band, all agreeing well with the data. The agreement (to within 20\%) is quite striking, considering the enormous complexity of the radiative transfer in the UV. The relative brightening with time (up to the on-set of gamma-ray leakage at $\sim$400 days) is due to decreasing line-blocking.

In the $B$-band, the models are also similar, reproducing the general time-evolution well. As in the $U$-band, the relative flux initially increases as line blocking decreases in efficiency. The small differences between the models in both $U$ and $B$ can be understood as a consequence of most of the radiation here originating from far out in the envelope (which has a similar structure in all models), the core being highly opaque due to line blocking.

From the $V$-band and on, line blocking is less efficient, and differences due to the different cores start to emerge. In $V$, the 12 and 15 \msun\ models are in good agreement with the observations at all times, whereas the 19 \msun\ model is too bright. The $V$-band includes \ion{Na}{i} \wll5890, 5896 and [\ion{O}{i}] \wll6300, 6364, and it is mainly too strong emission in these that cause the overproduction in the high-mass model (Sect. \ref{sec:spectra}).

The $R$-band behaves similarly to the $V$-band. The band is dominated by H$\alpha$, but is also influenced by [\ion{O}{i}] \wll6300, 6364, especially at late times. The fluxes in both lines are too strong in the 19 \msun~model (see below), overproducing the $R$-band. The 12 and 15 \msun\ models are better, but still show some discrepancies as they also over-produce H$\alpha$ (Sect. \ref{sec:spectra}). 

The $I$-band shows significant differences between the observed photometry and the synthetic photometry from the observed spectra, up to half a magnitude during the first year. The $I$-band is influenced by the [\ion{Ca}{ii}] 7291, 7323 doublet as well as the \ion{Ca}{ii} IR triplet, both lying on the edge of the filter. As mentioned before, this implies that SN $I$-band photometry has to be used with considerable caution.

\begin{figure*}[htb]
\centering
\includegraphics[trim=50mm 0mm 50mm 0mm, clip, width=0.95\linewidth]{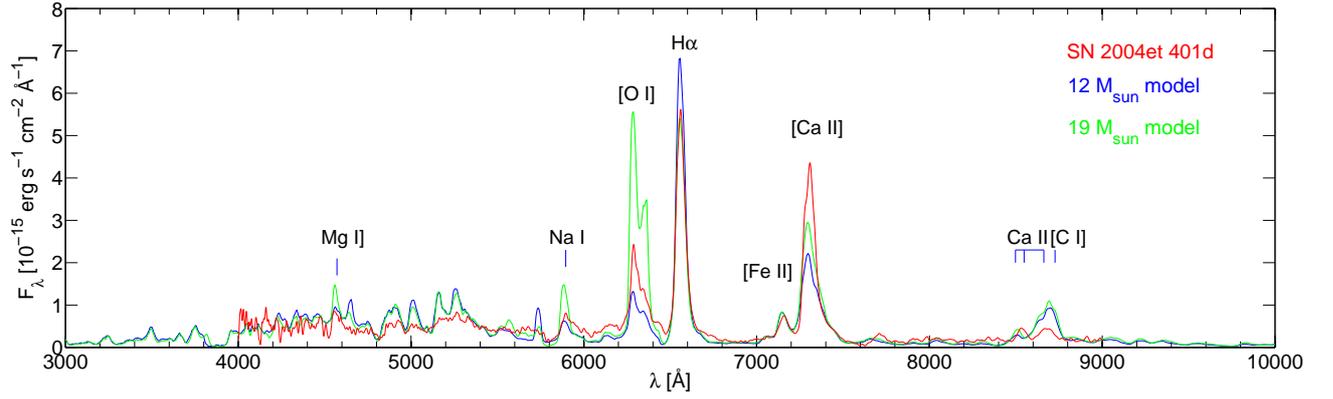}
\caption{Optical (dereddened) spectrum at 401 days (red) compared to the 12 \msun~(blue) and 19 \msun~(green) models.}
\label{fig:optical}
\end{figure*}

\begin{figure*}[htb]
\centering
\includegraphics[trim= 29mm 7mm 55mm 10mm, clip, width=0.95\linewidth]{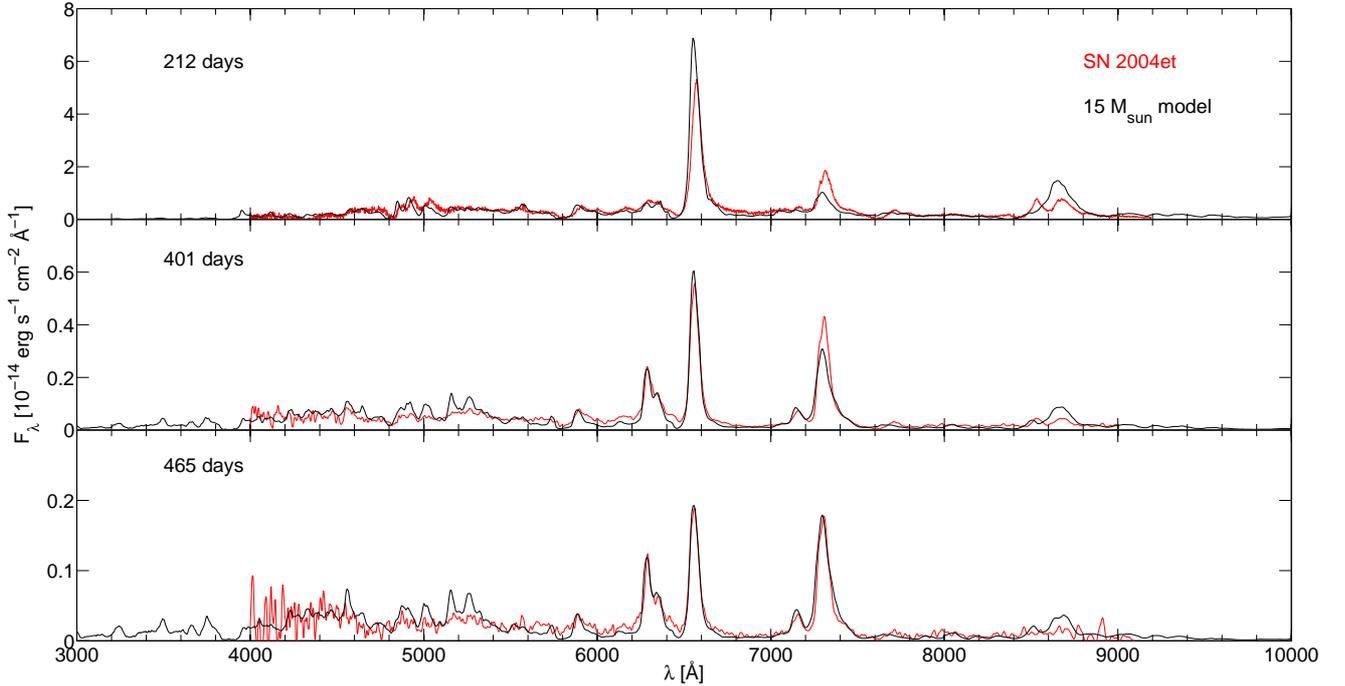}
\caption{Observed (dereddened) spectra (red) and 15 \msun~model spectra at 212, 401, and 465 days (black). All spectra have been smoothed with a Gaussian of FWHM = 600 \kms to improve the S/N.}
\label{fig:fig33}
\end{figure*}

In the NIR, the $J$ and $H$ data points at 134 and 141 days are important in showing that the models reproduce the flux reasonably well also in the NIR in the full gamma-ray trapping phase, which together with the good agreement in the other bands provide confirmation that the \iso{56}Ni mass is well determined. For the $K$-band, one should be aware that our models do not include the CO first overtone at 2.3 $\mu$m (which lies at the edge of the band), so we expect the models to be on the low side. 

It is clear that the progenitor mass has its strongest influence on the $V$, $R$, and $H$ bands, which are therefore valuable observables in the nebular phase. In $U$, $B$, $I$, $J$, and $K$, the differences are smaller.

\subsection{Spectral evolution}
\label{sec:spectra}

\subsubsection{Optical}
\label{sec:optspec}
Fig. \ref{fig:optical} shows the optical model spectra for the 12 and 19 \msun \ models compared to the observed (dereddened) spectrum at 401 days. It is clear that the main influence of the progenitor mass are the \ion{Na}{i} \wll5890, 5896 and [\ion{O}{i}] \wll6300, 6364 lines. The observed spectrum shows fluxes in these lines that fall in between the 12 \msun~and 19 \msun~models. 

Also H$\alpha$, [\ion{Ca}{ii}] \wll7291, 7323, and the \ion{Ca}{ii} IR triplet fluxes differ between the models, although by less. These lines are mainly formed in the hydrogen zone, and are therefore not direct indicators of the core mass, but rather depend on the quite uncertain pre-SN mass loss and the explosive mixing. 

The spectrum below 6000 \AA~is a complex mix of mainly iron-group lines formed by scattering and fluorescence in the hydrogen envelope. Because of the similar envelopes in the models, the model spectra differ by little in this spectral region. One exception is the \ion{Mg}{i}] \wl4571 line, which arises from the metal core. 




\begin{figure*}[htb]
\centering
\includegraphics[trim= 10mm 9mm 10mm 0mm, clip, width=0.9\linewidth,height=11cm]{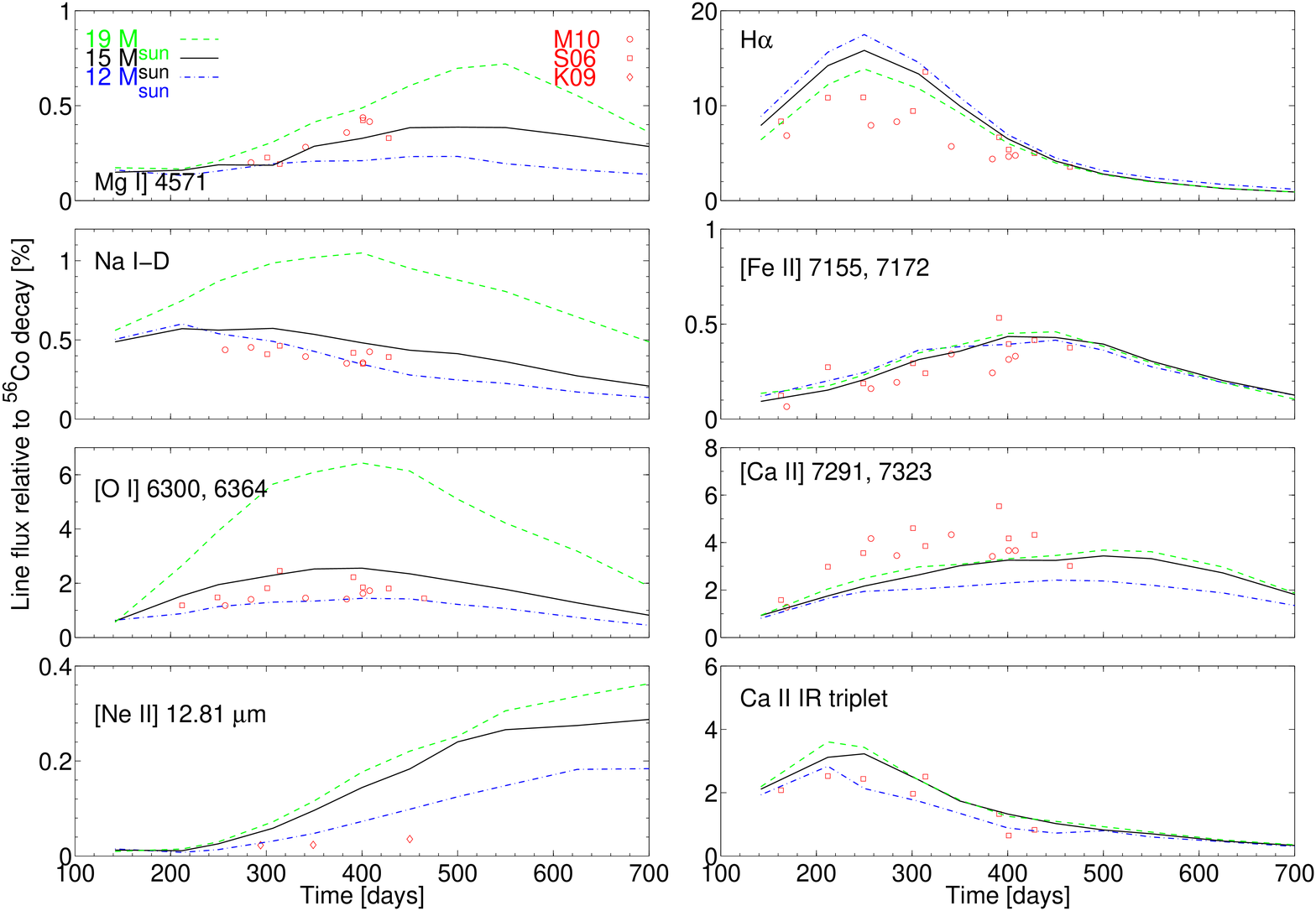}
\caption{Observed (dereddened) line fluxes (red points), normalized to the bolometric flux of 0.062 \msun~of \iso{56}Co at a distance of 5.5 Mpc, compared to model values (12 \msun\ : black, solid, 15 \msun\ : blue, dot-dashed, 19 \msun\ : green, dashed). See the text (Sect. \ref{sec:optspec}) for details on the line flux measurements.}
\label{fig:fig9}
\end{figure*}

\begin{figure*}[hbt]
\centering
\includegraphics[trim = 55mm 10mm 55mm 5mm, clip=true, width=1\linewidth]{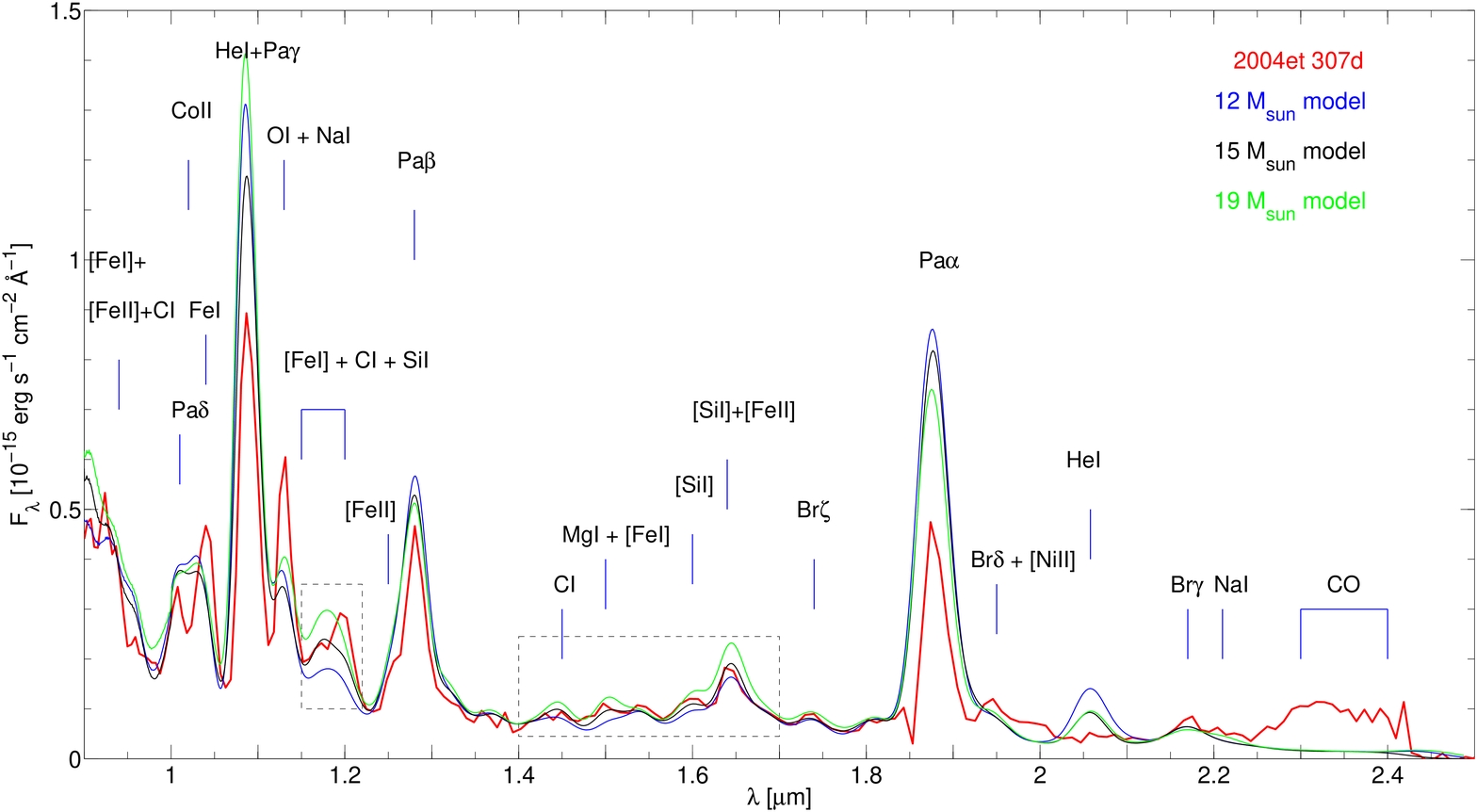}
\caption{Near-infrared (dereddened) spectrum at 307 days (red) compared to the 12 \msun~(blue), 15 \msun~(black), and 19 \msun~(green) models. The boxed regions at $1.5-1.2$ $\mu$m and $1.4-1.7$ $\mu$m show particular sensitivity to the core mass. Note that CO emission is not included in the model spectra.}
\label{fig:NIR}
\end{figure*}

The 15 \msun~model spectra at 212, 401, and 465 days are plotted in Fig. \ref{fig:fig33}. We see that this model is generally successful in reproducing the observed spectra, and in particular reproduces the [\ion{O}{i}] \wll6300, 6364 lines well at all epochs. This is further corroborated by Fig. \ref{fig:fig9}, where we plot the evolution of \ion{Mg}{i}] \wl 4571, \ion{Na}{i} \wll 5890, 5896, [\ion{O}{i}] \wll6300, 6364, [\ion{Ne}{ii}] \wl12.81 $\mu$m, H$\alpha$, [\ion{Fe}{ii}] \wll7155, 7172, [\ion{Ca}{ii}] \wll 7291, 7323, and the \ion{Ca}{ii} IR triplet, from the observed spectra as well as from the three different models. Both observed and modeled fluxes have been normalized to the bolometric flux of 0.062 \msun~of \iso{56}Co, assuming a distance of 5.5 Mpc. The resulting numbers therefore represent the fraction of the radioactive decay energy re-emerging in the various lines. The optical line flux measurements were done by integrating the spectra over $V_{\rm cut}=\pm 2700$ \kms~and subtracting the continuum level, which was taken as the linear interpolation of the minimum flux values on the right and lefthand side, within $V_{\rm cont}=\pm 1.25 V_{\rm cut}=\pm3375$ \kms. These velocity limits were found to give reasonable extractions of the line fluxes for all lines and epochs, except for H$\alpha$ for which we used $V_{\rm cut}=\pm7200$ \kms \ and $V_{\rm cont}=1.25 V_{\rm cut} = \pm9000$ \kms\ to include the broad wings. Because we perform the same operation on model and observed spectra, the choice is not critical for the comparison.

\paragraph{Diagnostic lines of the core mass}
\label{sec_core}
The most important lines as diagnostics of the core mass are [\ion{O}{i}] \wll6300, 6364. These lines are distinct at most epochs, relatively undisturbed by line-blending and line blocking (there is, however, some iron line emission blended with them, see \citet{Dessart2011}), and have at least half their flux coming from synthesized oxygen (M12). In addition, oxygen is mainly neutral in the core, while magnesium and sodium are mainly ionized, meaning that the thermal emission by \ion{O}{i} is quite insensitive to the ionization balance. In all models, the thermal emission by [\ion{O}{i}] \wll6300, 6364 from the O/Ne/Mg zone corresponds to $50-60 \%$ of the heating of this zone, at all times after 300 days. Combined with the constraints on the density that the 6300/6364 flux ratio provides (M12), the doublet lines are excellent diagnostics of the core mass. The main complication is that primordial oxygen in the hydrogen zone contributes to the emission as well (M12). In the 15 and 19 \msun~models, however, there is so much synthesized oxygen that this contribution is quite small.

Also the \ion{Mg}{i}] \wl4571 line is a potentially useful diagnostic of the O/Ne/Mg zone. According to the models, this is the strongest optical \ion{Mg}{i} line, followed by \ion{Mg}{i} \wll5167, 5172, 5183, which is too weak to be detected. 
The \ion{Mg}{i}] \wl4571 line is, however, also rather weak and difficult to extract. The first clear detection is in the spectrum taken at 284 days. Before that, complex line blending and blocking makes it difficult to distinguish, and even later it is possible that the emission is blended. The last spectrum at day 465 has too poor S/N to assess the line. 

None of the models produce any distinct \ion{Mg}{i}] \wl4571 until $200-300$ days after explosion, in agreement with the observations. The evolution after that is in best agreement with the 15 \msun~model, but the model differences are quite small. Observations after 500 days would have allowed a more conclusive comparison between the models to have been made with regard to this line.

\ion{Mg}{ii} is the dominant ion in the O/Ne/Mg zone at all epochs, so the \ion{Mg}{i} lines have strong contributions by recombination. The transition has a thermal emission of about 10\% of the [\ion{O}{i}] \wll6300, 6364 emission. This is not enough to account for the luminosity of the line (Fig. \ref{fig:fig9}), so recombinations indeed dominate the emission.


The \ion{Na}{i} D resonance lines are optically thick quite far out in the H-envelope (to 6200 $\kms$ at day 300). The lines are therefore formed by a combination of P-Cygni scattering and by emission by synthesized sodium in the core. The D lines are responsible for some of the cooling of the O/Ne/Mg zone, but most of the emission is recombination. The observed flux is in good agreement with the 12 and 15 \msun~models, but is strongly overproduced by the 19 \msun~model. Less influenced by line blocking and blending than the \ion{Mg}{i}] \wl4571 line, but still sensitive to charge transfer (J11) as well as the hydrogen envelope properties (through the scattering), this line is intermediate to the \ion{Mg}{i}] \wl4571 and [\ion{O}{i}] \wll6300, 6364 lines in model parameter sensitivity.

We do not reproduce the [\ion{Ne}{ii}] 12.81 $\mu$m line well with any of the models. We discuss this line more in the section about the MIR lines.

\paragraph{Diagnostic lines of the hydrogen and helium zones}

In addition to the the hydrogen and helium lines, also many metal lines are dominated by emission from the hydrogen and helium zones. The strongest lines from these zones are discussed in detail in M12, and we will therefore here only make some specific points relevant for this paper. 

The hydrogen envelope mass is similar in the three models,  7.7 - 9.5 \msun~(Table \ref{table:zonemasses}). We also assume the fraction mixed into the core to be the same in all models ($x_{\rm H}=0.15$).  There are therefore no major differences in the H-zone lines between the models. Still, as Fig. \ref{fig:fig9} shows, the lines from this zone do differ somewhat, as the hydrogen and helium zones have different densities for their in-mixed components, and are also exposed to different amounts of gamma-rays as well as diffuse radiation, depending on the metal content of the core.

H$\alpha$ is formed by recombinations following photoionizations in the Balmer continuum up to $\sim$500 days, and recombinations following non-thermal ionizations from the ground state after that. With less core material to absorb the gamma-rays in the lower-mass models, the H$\alpha$ flux is somewhat higher, although the hydrogen mass is smaller. The observed flux is lower than in all models. As most of the model flux comes from the in-mixed hydrogen component, a reduction in in-mixing below 15\% would likely improve the fits. 
Note, however, that the H$\alpha$ flux in SN 1987A was about a factor of two weaker than expected (from the other hydrogen recombination lines) in the $200-500$ day range  \citep{Xu1992}, still without an explanation.

The [\ion{Ca}{ii}] \wll7291, 7323 lines are significant coolants of the hydrogen zone \citep[][KF98b, M12]{Li1993}. The models differ by little, both agreeing reasonably well with the data. The ratio of the \ion{Ca}{ii} triplet lines to the 7291, 7323 \AA\ lines depends on the temperature and density \citep{Fransson1989,Ferland1989,Li1993}. Our models give a slightly too high triplet to 7291, 7323 \AA\ ratio, which indicates that our density in the \ion{Ca}{ii} emitting gas may be too high.

\subsubsection{Near-infrared}

Fig. \ref{fig:NIR} shows the day 307 NIR spectrum compared to the models. It is clear that the core mass has a significant influence on the spectral features at $1.15-1.2$ $\mu$m and $1.4-1.7~\mu$m. The plateau-like feature at $1.15-1.2$ $\mu$m is a blend of \ion{C}{i} \wl1.176, 1.181 $\mu$m (C/O zone + He/C zone), \ion{Mg}{i} \wl1.183 $\mu$m (O/Ne/Mg zone), \ion{Si}{i} \wl1.200 $\mu$m (Si/S zone + O/Si/S zone), as well as several \ion{Fe}{i} lines from both the Fe/He zone and from the O/Ne/Mg zone. The $1.4-1.7~\mu$m range contains emission by \ion{C}{i} \wl1.454 $\mu$m (C/O zone + He/C zone), \ion{Mg}{i} \wl1.504 $\mu$m (O/Ne/Mg zone), [\ion{Si}{i}] \wl1.607, 1.645 $\mu$m (Si/S zone), and [\ion{Fe}{ii}] \wl1.534 $\mu$m, 1.600 $\mu$m, 1.644 $\mu$m, 1.681 $\mu$m (Fe/He zone). As an important complement to the optical analysis, we find that  the 15 \msun~model produces better fits than the 12 and 19 \msun~models for these metal emission lines as well. 

The \ion{O}{i} + \ion{Na}{i} feature at 1.13 $\mu$m may seem to suggest a higher-mass progenitor than 19 \msun. However, the \ion{O}{i} \wl1.129 $\mu$m line (which dominates the blend) is produced by \ion{H}{i} Ly$\beta$ branching into \ion{O}{i} \wl1025.76 $2p^4(^3P) - 3d(^3D^o)$, and \citet{Oliva1993} found that much of this branching must occur over zone boundaries, with Ly$\beta$ photons scattering into regions of synthesized oxygen while trapped between $n=1$ and $n=3$ in \ion{H}{i}. Our modeling does not treat such line overlap between zones (although we do treat the overlap \emph{within} each  zone, see App. \ref{sec:lineoverlap}), and we can therefore not properly model the line.

The strongest feature in the NIR is \ion{He}{i} \wl1.083 $\mu$m + [\ion{S}{i}] \wl1.082 $\mu$m, which is a bit too strong in the models. 
Of the hydrogen lines, both Pa$\alpha$ and Pa$\beta$ are also overproduced. Together with the overproduction of H$\alpha$, this suggests that the He and H content of the models may be too large, and/or too strongly mixed into the radioactive core.

\subsubsection{Mid-infrared}
\label{sec:mir}

As shown by K09, and discussed in Sect. \ref{sec_mod_par}, the mid-IR is dominated by a combination of warm dust emission from the ejecta and cold dust emission from the circumstellar environment, with superimposed atomic and molecular emission lines from the ejecta. To reproduce the spectrum in this region it is therefore necessary to include these dust components. For this we repeat the modeling in K09, although with slightly different parameters, since we here include also the detailed contribution by atomic emission (which was approximated as a hot blackbody in K09). 

For ejecta dust extinction we use the smoothed Draine \& Lee model with ``astronomical'' silicate \citep{Draine1984,Laor1993} and use the escape probability formalism (Eq. (\ref{eq:escapeprob})). As parameters at 350 and 451 days we find temperatures of 700 and 600 K, and optical depths at 10 $\mu$m of 2.1 and 3.2, respectively, to be good choices for reproducing the observed spectrum. These values are similar to the ones derived by K09. For the circumstellar dust component, we assume a blackbody SED, and obtain a good fit at 350 days with $T = 50$ K and $L= 3\e{38}$ \ergs.  

 We also add a component of vibrational-rotational SiO lines from the fundamental band at $7-9$ $\mu$m. These are calculated from a NLTE solution of the SiO molecule, including the first 100 rotational levels of the $v=0$ to $v=3$ vibrational levels, with collision rates from \citet{Varambhia2009} and \citet{Liu1994}, and radiative transition rates from \cite{Drira1997}. 
 The parameters for this component are the temperature, the electron density, and the total luminosity. For the electron density we take $n_{\rm e} = 7\e7 \ \ccm$ at 350 days and $n_{\rm e} = 2.6\e7 \ \ccm$ at 451 days, as given by the models for the O/Si/S zone. Because of the limited S/N and resolution of the spectra, the shape of the P and R branches of the $v = 1$ to $v=0$ bands are not well defined. We therefore use a temperature similar to that in SN 1987A of 1500 K \citep{Liu1994}. We assume a velocity equal to that of the core velocity. For the luminosity, we find that $L_{\rm SiO} = 9.9\e{37} \ergs$ at 350 days and $L_{\rm SiO} = 4.3\e{37} \ergs$ at 451 days reproduce the observed spectra in the $7-9$ $\mu$m range well.

\begin{figure*}[htb]
\centering
\includegraphics[trim=52mm 5mm 55mm 5mm, clip, width=1\linewidth]{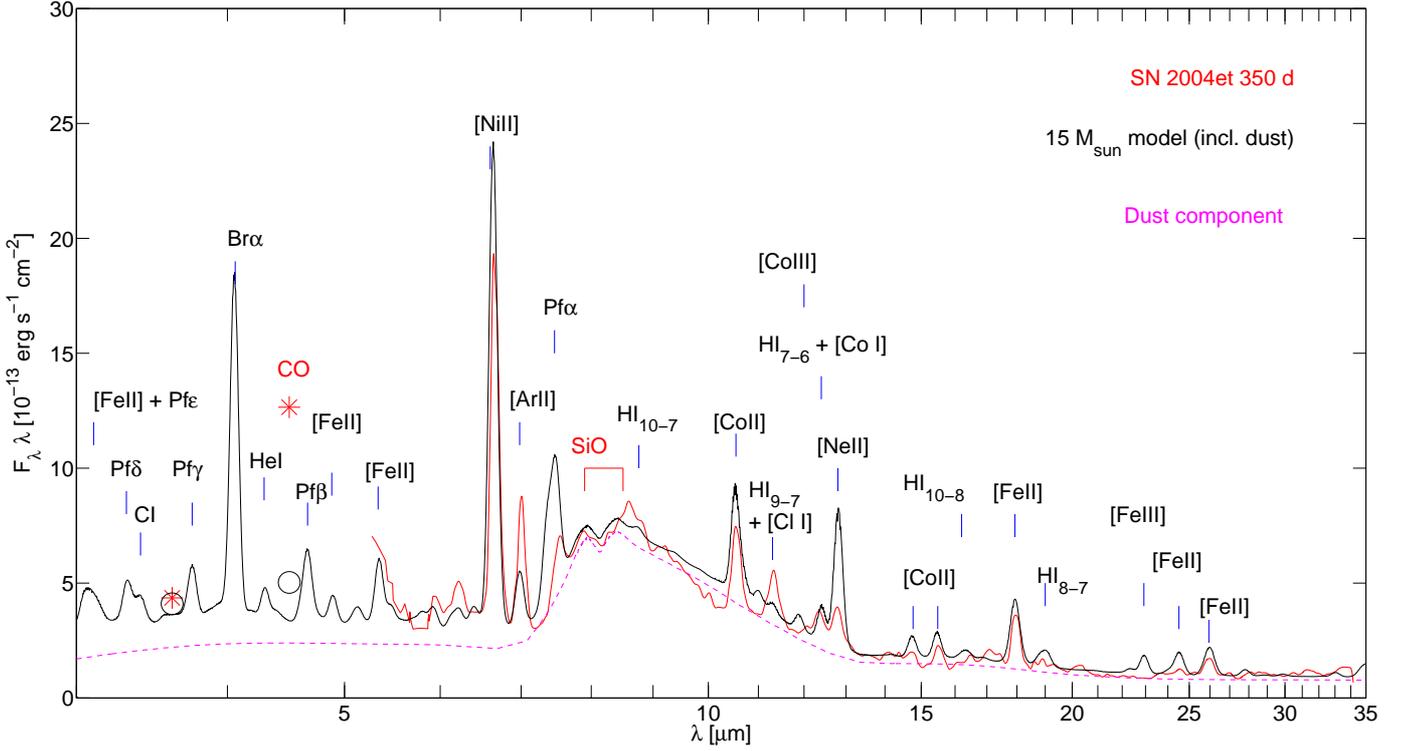}
\caption{Observed MIR spectrum at 350 days (red, solid line), compared to model spectra of the 15 \msun~model (black, solid). To simulate the instrument response, the model has been convolved with a Gaussian of $FWHM=\lambda/R$, where the resolving power $R$ was taken to be 60. Also shown are the dust component (pink, dashed), the observed photometry at 3.6 and 4.5 $\mu$m (red stars), and the corresponding model photometry for the model (black circles). The observed photometry is at 361 days, and has been rescaled with the \iso{56}Co decay factor to 350 days ($\exp{-11/111.5}$).}
\label{fig:fir350_2}
\end{figure*}


Figs. \ref{fig:fir350_2} shows the Spitzer spectrum at 350 days (red line), together with the 15 \msun~model spectrum including the ejecta dust, circumstellar dust, and SiO components described above (black line). It is clear that above 8 $\mu$m, dust emission dominates the atomic emission. At shorter wavelengths, however, the atomic emission is about as strong as the dust emission. It is also clear that CO $4.55~\mu$m fundamental band emission is present, as there is not enough emission by dust and atoms in the $4-5$ $\mu$m range to account for the photometric 4.5 $\mu$m observations. To reach the observed flux, we need $L_{\rm CO}=6.0\e{38}$ \ergs~at 350 days, and for 295 and 451 days we need $L_{\rm CO}=4.1\e{38}$ \ergs \ and $L_{\rm CO}=2.1\e{38}$ \ergs, respectively. That significant amounts of CO have formed is also evidenced by the presence of the first overtone band at $2.3~\mu$m (Fig. \ref{fig:NIR}). The presence of both CO and SiO emission supports our assumption of molecular cooling occurring in the O/Si/S and O/C zones (see also Sec. \ref{sec:molecules}).

The atomic lines we identify using our model spectra are labeled in Fig. \ref{fig:fir350_2}. Most lines do not differ significantly between the models, as they mainly originate in the Fe/He or H zones. The only line sensitive to the core mass is [\ion{Ne}{ii}] \wl12.81 $\mu$m. This line, arising in our models mainly from the He/C zone, but also from the O/Ne/Mg zone, is too strong in all models, suggesting that the neon content in this zone is too high, that we have mixed the He/C zone too strongly into the radioactive core, or that our ionization solution overproduces singly ionized neon. 

In our models, the [\ion{Ar}{ii}] \wl6.985 $\mu$m line is not emitted by synthesized argon, but by primordial argon in the He zone. Here, argon is singly ionized and is responsible for about 1\% of the cooling. The synthesized argon is located mainly in the Si/S zone, but is here only ionized to $\sim10^{-4}-10^{-3}$, and provides only $\sim 10^{-4}$ of the cooling, which is dominated by [\ion{S}{i}] and [\ion{Ca}{ii}].

\subsection{Determination of the iron zone filling factor}
\label{sec:izff}
As demonstrated for SN 1987A, the infrared iron-group lines are useful as diagnostics of the physical conditions in the \iso{56}Ni ``bubble'', resulting from radioactive \iso{56}Ni heating \citep{Moseley1989,Haas1990,Spyromilio1992,Roche1993,Li1993iron}. A complication is, however, that there are varying contributions to these lines from primordial iron from the H and He zones (KF98b, M12). In addition, the studies mentioned make several assumptions about the ionization balance and optical depths. More reliable results require calculation of these quantities. 

Because both the Fe/He and H/He zones are similar in the various models, the iron-group lines differ little between them, and we consider only the 15 \msun~model. The electron density in the Fe/He zone at 350 days is of order $10^9 \ccm$ in the model, much higher than the critical densities for forbidden fine-structure lines ($\sim$$10^5 \ccm$), so all lines arising from the lower multiplets are in LTE. 
We find, for all possible filling factors of the iron clumps, optical depths much higher than unity for the [\ion{Fe}{ii}] \wl17.94 $\mu$m and [\ion{Fe}{ii}] \wl25.99 $\mu$m lines. These lines are thus both in LTE and optically thick. The general expression for line luminosity, in the Sobolev approximation, is
\begin{equation}
L(t) = V(t) n_{\rm u}(t) A \frac{1-e^{-\tau_s}}{\tau_s}h\nu
\label{eq:sob}
\end{equation}
where $V(t)$ is the zone volume, $n_{\rm u}(t)$ is the number density of the upper level, $A$ is the transition probability, and $\tau_{\rm s}$ is the Sobolev optical depth.
For an optically thick line in LTE, Eq. (\ref{eq:sob}) becomes
\begin{equation}
L(t) = \frac{V(t)}{t} \frac{8\pi h\nu}{\lambda^3\left(e^{h\nu/kT(t)}-1\right)}
\end{equation}
Since $h\nu \ll kT$ for the MIR lines, the line luminosity is then
\begin{equation}
L(t)  \propto  \lambda^{-3}f t^2 T(t)~\ ,
\label{Eq:lltlte}
\end{equation}
where $f$ is the filling factor. Since we, at a given time $t$, can compute the temperature $T(t)$ for any given filling factor $f$, we are therefore in a good position to determine the latter by comparing model line luminosities to the observed ones.

\begin{figure}[htb]
\includegraphics[clip,trim = 5mm 0mm 5mm 0mm, width=1\linewidth]{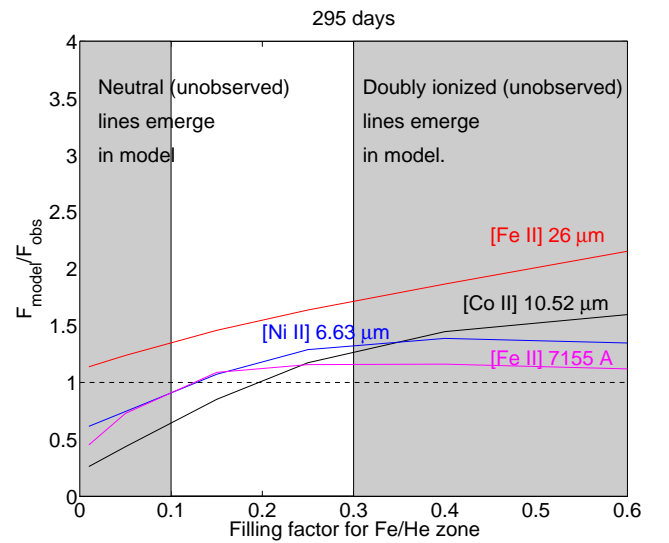}  
\caption{Model Fe/Ni/Co line fluxes relative to observed values, at 295 days. The 15 \msun~model has been used.}
\label{fig:fFeHe_295}
\end{figure}

\begin{figure}[htb]
\includegraphics[width=1\linewidth,clip,trim = 5mm 0mm 5mm 0mm]{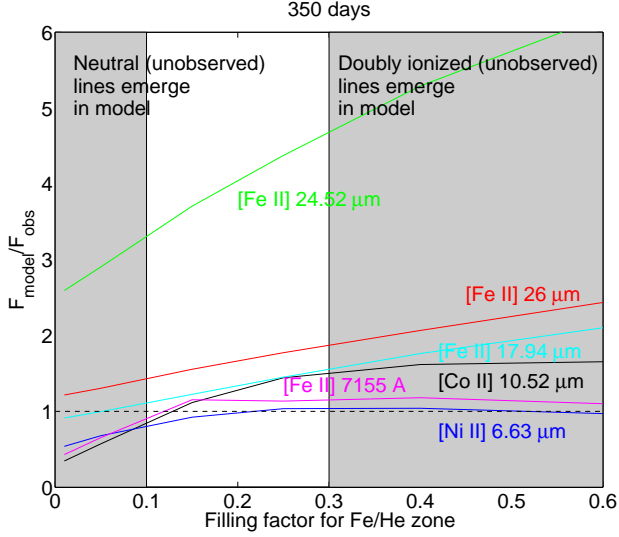}
\caption{Same as Fig. \ref{fig:fFeHe_295}, for 350 days.}
\end{figure}

\begin{figure}[htb]
\includegraphics[width=1\linewidth, clip,trim = 5mm 0mm 5mm 0mm]{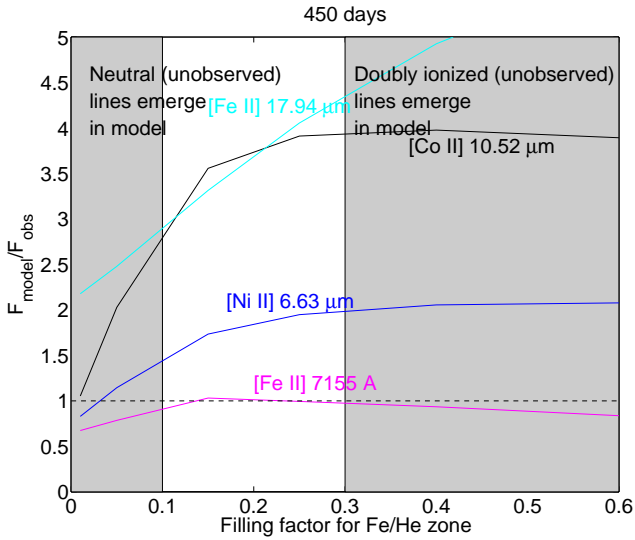}
\caption{Same as Fig. \ref{fig:fFeHe_295}, for 451 days.}
\label{fig:fFeHe_451}
\end{figure}

Figs. \ref{fig:fFeHe_295}-\ref{fig:fFeHe_451} show the model fluxes of the Fe/Co/Ni MIR lines relative to the measured fluxes, as a function of $f_{\rm Fe/He}$. When we vary $f_{\rm Fe/He}$, we add/subtract the corresponding amount to $f_{\rm H}$. We measure the line fluxes as described in Sect. \ref{sec:optspec}, using $V_{\rm cut}=\pm6400$ \kms($=\pm 3\sigma$ for $R=60$), and $V_{\rm cont}=1.25 V_{\rm cut}=\pm8000$ \kms. The model spectra were convolved to the Spitzer resolution ($R=60$), before this operation. 
For each epoch we only plot the lines with clear detections. 

The MIR lines behave quite consistently, many of them being optically thick and in LTE. The filling factor-dependency is generally weaker than Eq. (\ref{Eq:lltlte}) suggests, as there is also an (almost constant) component from the He/H zones. We find that we can fit the Fe/Ni/Co lines only if $f_{\rm Fe/He}=0.1-0.2$. At larger values, the singly ionized lines become too strong, and in addition doubly ionized lines  ([\ion{Fe}{iii}] \wl22.94 $\mu$m, [\ion{Co}{iii}] \wl11.89 $\mu$m, [\ion{Co}{iii}] \wl16.39 $\mu$m) emerge in the model. These lines are not observed, and we therefore rule out a filling factor larger than $f_{\rm Fe/He} \approx 0.3$. Values much smaller than $f_{\rm Fe/He} \approx 0.1$ can also be ruled out, as the singly ionized lines are then too weak, and unobserved neutral lines in the optical/NIR (\ion{Fe}{i}] $4s(a^5F)-4s4p(z^7D^o)$ transitions between $8000-8500$ \AA\ and [\ion{Fe}{i}] $4s^2(a^5D)-4s(a^5F)$ transitions at 1.36, 1.44 and 1.53 $\mu$m), are strong. A filling  factor of $f_{\rm Fe/He}=0.1-0.2$ is also supported by the [\ion{Fe}{ii}] \wl7155 line, whose behaviour agrees well with the MIR lines, removing any uncertainty regarding the MIR calibration for our conclusions. 

As is seen in the figures, [\ion{Fe}{ii}] \wl24.52 $\mu$m is not reproduced well by any filling factor. This line is, however, weak and in a noisy part of the spectrum. Also the [\ion{Fe}{ii}] \wl25.99 $\mu$m line shows poorer agreement compared to the shorter wavelength lines.

Our result here is close to the $f_{\rm Fe/He}$ value derived for SN 1987A by KF98b ($f_{\rm Fe/He}\approx 0.2$). The value derived by \citet{Li1993iron} ($f_{\rm Fe/He} > 0.3$) is likely too large as they ignore the primordial Fe emission. We find that for $f_{\rm Fe/He}=0.15$ and $t=350$ days, the contribution by primordial iron-group elements in the H and He zones to the line fluxes is $30-40$\%  for the [\ion{Fe}{ii}] lines, and 20\% for [\ion{Ni}{ii}] \wl6.636 $\mu$m. The contribution by primordial cobalt is low ($<1\%$) as the solar abundance of cobalt is very low. The [\ion{Co}{ii}] \wl10.52 $\mu$m line thus comes from the radioactive \iso{56}Co, but being optically thick (up to 500 days) does not allow an independent determination of the \iso{56}Co mass. Also the [\ion{Co}{ii}] \wl14.74 and 15.46 $\mu$m lines have significant optical depth ($\tau \approx$ 1 at 350 days).

The [\ion{Ni}{ii}] \wl6.634 $\mu$m line arises mainly from stable  \iso{58}Ni in the Fe/He zone, with some contribution from  \iso{57}Ni. We find that the [\ion{Ni}{ii}] 6.634 line is well produced by the amount of \iso{58}Ni in the explosion models ($M(^{58}\mbox{Ni})/M(^{56}\mbox{Ni})\approx 0.04$). 

\subsubsection{Molecules}

\begin{figure}[htbp]
\centering
\includegraphics[trim=5mm 0mm 5mm 0mm, clip, width=1\linewidth]{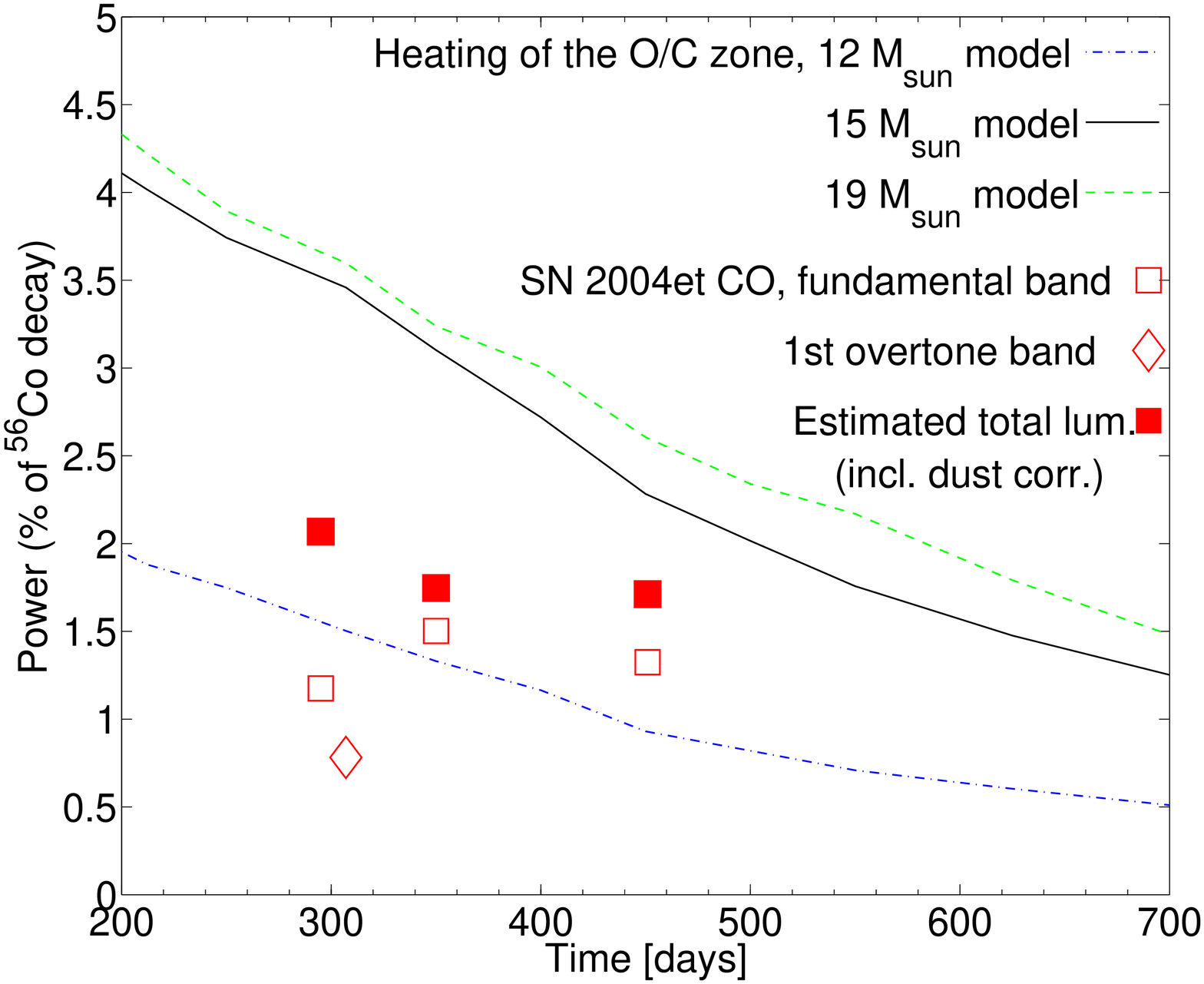}
\caption{Observed flux in the CO fundamental band (hollow squares), first overtone (diamond), and estimated total (filled squares), compared to the total heating of the O/C zone in the various models. For the estimated total emission, correction for the internal dust extinction has been taken into account. All values have been normalized to the decay of 0.062 \msun~of \iso{56}Co.}
\label{fig:mols}
\end{figure}

\begin{figure}[htbp]
\centering
\includegraphics[trim=5mm 0mm 5mm 0mm, clip,width=1\linewidth]{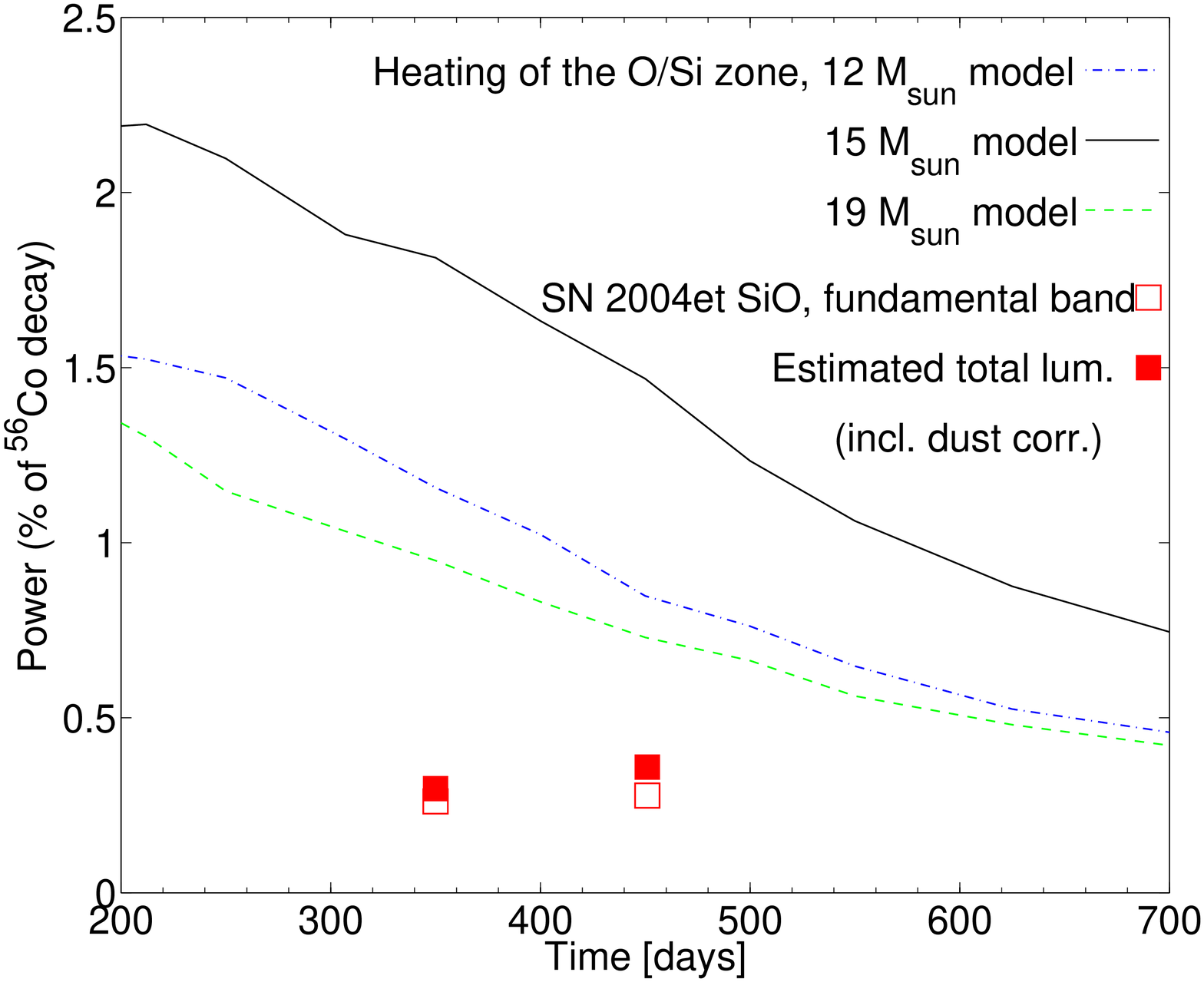}
\caption{Same as Fig. \ref{fig:mols}, for SiO. The overtone band is unobserved, and is assumed to be zero.}
\label{fig:molsio}
\end{figure}

Figs. \ref{fig:mols}-\ref{fig:molsio} show the estimated total luminosities of CO and SiO (Sect. \ref{sec:mir}), compared to the heating of the O/C and O/Si/S zones in the three models. If molecules are responsible for all of the cooling in these zones, these values should match for the model with the correct O/Si/S and O/C zone masses.

For SiO, no observations of the overtone exist, and for CO, we have only one observation. The CO overtone luminosity at the other epochs is assumed to scale to the day 307 luminosity in the same manner as in SN 1987A \citep{Bouchet1993}. It contributes less than 10\% to the total luminosity at 350 and 450 days.

\label{sec:molecules}


The CO emission falls in between the 12 and 15 \msun\ models, behaving consistently with the optical emission lines. This result is also important in ruling out CO as a possibility for a more massive progenitor to still produce weak [\ion{O}{i}] 6300, 6364 \AA\ emission from the O/Ne/Mg zone. The observed CO flux in such a case would be much stronger.

The SiO emission is about a factor of 2-3 below the total heating of the O/Si/S zone in all of the models. As for CO, this is important for constraining the possible cooling channels of the O/Ne/Mg zone. Note that O/Si/S zone does not grow monotonically with progenitor mass, being larger in the 15 \msun~model than in the 19 \msun~one (Table \ref{table:zonemasses}).

It is possible that SiO only does part of the cooling of the O/Si/S zone. However, with the large partition function of molecules, one would expect it to dominate the cooling. 
The determination of the SiO fundamental band flux is complicated by the silicate dust feature at the same wavelength, possibly explaining the discrepancy. It is also likely that the silicate dust, which presumably forms in the same zone as the SiO, produces stronger extinction there than for the spectrum as a whole, and/or does some of the cooling.

\section{Discussion}
\label{sec_disc}

In evaluating the robustness of linking the nebular spectra to a progenitor mass, there are several aspects to consider. Our models test specific stellar evolution/explosion models (in our case the models by WH07), and the zone masses and compositions in these may vary depending on the assumptions about convection, rotation, and nuclear burning rates \citep[see e.g.,][]{Schaller1992, WW95, Thielemann1996, Woosley2002, Hirschi2004}. 
It is beyond the scope of this paper to test all types of models with respect to variation in these parameters. However, an attempt to judge the influence of some of them is warranted. 

The influence of metallicity on the total nucleosynthesis yield is quite weak, although it does affect specific elements in particular zones  (WW95). Estimates of the metallicity in the region of SN 2004et indicate that it is likely within a factor two from solar (Sect. \ref{sec:metallicity}), and such a variation should not have any strong impact on the stellar evolution or SN spectral formation. Although it was found by M12 that emission by primordial metals in the hydrogen envelope can be significant contributors to the nebular spectrum, this emission will change little with metallicity as long as the \emph{relative} amounts of metals are roughly the same. The fraction of cooling done by each element will then roughly stay the same, although a slight temperature increase will occur for lower metallicities due to the fewer number of coolants available. We therefore do not think that the assumption of solar metallicity is crucial for any of our conclusions.

Rotation can enhance the nucleosynthesis yield for a given stellar mass \citep[e.g.][]{Hirschi2004}. For rapidly rotating progenitor models, our best fit would (most likely) have been a lower-mass star. With regard to this parameter, we can therefore say that 15 \msun~is an upper limit to the progenitor mass. One should note, however, that the progenitor analyses \citep[e.g.][]{Crockett2011} also assume non-rotating stars, so if one postulates rotation, best-fitting progenitor models also go down in mass (rotation boosts luminosity in the late evolutionary stages).

Assumptions about the symmetry, mixing, and fragmentation of the explosion also influence the model outputs. The observed line profiles constrain the velocity range of the material, however, and furthermore do not indicate any extreme asymmetries in the SN. As gamma-rays scatter around the whole ejecta in the nebular phase, neither is it obvious that even strong asymmetries will significantly change the energy deposition in the various zones. 

Our results here are in fair agreement with the progenitor analysis of SN 2004et, which has given $M_{\rm ZAMS}=11_{-1}^{+2}$ \msun~(Fraser et al. 2012, in prep.). On the other hand, hydrodynamical modeling by \citet{Utrobin2009} gave an ejecta mass of 24.5 \msun, over a factor two larger than the ejecta mass of our best-fitting model here (10.9 \msun), as well as any final mass in standard stellar evolution models (the maximum ejecta mass with standard mass-loss rates is $\sim$14 \msun \citep{Kasen2009}, and will be even lower if strong pulsation-driven mass-loss occurs in the RSG phase \citep{Yoon2010}). 
The discrepancy between the hydrodynamical modeling and the progenitor analysis/nebular-phase spectral analysis is important to understand. More independent analyses in all three fields are needed to make progress here. A possible concern regarding the hydrodynamical modeling by \citet{Utrobin2009} is that non-evolutionary progenitor models are used.

We also mention the results in \citet{Chevalier2006}, who estimated a progenitor mass of $\sim$20 \msun~by linking the estimated mass-loss rate from radio observations to standard mass-loss laws and stellar evolution models. This result relies on several assumptions (wind velocity, shell velocity, circumstellar temperature, and dominance of free-free opacity), and in addition the mass-loss laws have standard deviations of a factor $2-3$. \citet{Rho2007} derived a lower mass-loss rate of $2\e{-6}$ \msun\ yr$^{-1}$, which corresponds to a $M_{\rm ZAMS}=15$ \msun\ progenitor.

Comparing with other nebular-phase analysis of SN 2004et, we reach a similar conclusion as S06 and  M10 do by simply comparing the [\ion{O}{i}] doublet to SN 1987A, that the oxygen mass is somewhat lower than in SN 1987A, and the progenitor thereby a bit less massive (assuming similar rotation etc.). We have here, however, put this result on a much firmer footing by computing detailed models and finding consistent results from several other lines. One should also bear in mind the contribution by primordial oxygen that complicates such simple comparisons (M12). The comparison to SN 1987A works quite well in this case as these two SN both have large masses of synthesized oxygen, similar velocities, and similar \iso{56}Ni-masses. One should note though, that the progenitor mass of SN 1987A cannot be taken to be known better than in the range $16-22$ \msun\ \citep{Arnett1989}. Since modeling of SN 1987A would require only small changes to the parameters used here, we would most likely find a mass on the lower side of this range. 

\citet{Dessart2010} find, using models without any macroscopic mixing, a correlation between the width of the oxygen lines and the progenitor mass (for a fixed explosion energy). For an explosion energy of 2.3 B, they find that the progenitor mass of SN 2004et should be less than 18 \msun, which is in agreement with our result here.

For a more general comparison of our modeling with other work, we consider the results in \citet{Dessart2011}, who
compute model spectra for 15 and 25 \msun~progenitors, also taken from WH07. These authors include a full NLTE and radiative transfer treatment, and in this sense their work is complementary to ours. They do, however, assume some microscopic mixing to occur, which may have important effects on the cooling and ionization balance, and therefore also on the emerging spectrum. At 300 days, the difference in the flux of the [\ion{O}{i}] doublet appears to differ by the same amount as the \iso{56}Ni mass in the two models (factor 1.9), a result possibly in disagreement with our model, but since we use the same \iso{56}Ni mass for all models we cannot say with certainty. At 1000 days the more massive star shows about a factor of 3.6 stronger [\ion{O}{i}] \wll6300, 6364 emission, or a factor 1.9 stronger per unit \iso{56}Co mass.

In M12, it was shown that the diversity of line emission in a sample of Type IIP SNe is quite small. One reason for this was found to be that the large hydrogen envelopes in these objects add significant contribution to all metal emission lines, mitigating the impact of varying core masses. Despite this damping effect, our modeling here suggests that stars at and above 19 \msun\ should produce \ion{Na}{i} \wll5890, 5890, [\ion{O}{i}] \wll 6300, 6364, [\ion{Ne}{ii}] \wl12.81 $\mu$m, [\ion{Si}{i}] \wl1.60, 1.64 $\mu$m about a factor two stronger than lower-mass progenitors (15 \msun\ and below). None of the SNe in the M12 sample, including SN 1987A, show such strong oxygen emission. Future detailed modeling of individual objects (making sure the ``missing flux'' cannot be accounted for in other lines such as CO), and a larger sample size, will show if these high-mass stars are indeed missing from the Type IIP group.

\section{Conclusions}

\begin{itemize}
\item We have shown that nebular-phase spectral modeling can be used to constrain the progenitor masses of Type IIP supernovae. Lines showing significant dependency on the progenitor mass are [\ion{O}{i}] \wll6300, 6364,  \ion{Na}{i} \wll5890, 5896, \ion{Mg}{i}] \wl4571, \ion{Mg}{i} \wl1.503 $\mu$m, \ion{C}{i} \wl1.176 $\mu$m, \ion{C}{i} \wl1.454 $\mu$m, \ion{Si}{i} \wl1.20 $\mu$m, [\ion{Si}{i}] \wll1.607, 1.645 $\mu$m, and [\ion{Ne}{ii}] \wl12.81 $\mu$m. 
\item The nebular-phase optical and near-infrared emission lines of SN 2004et are well reproduced by model spectra of a $M_{\rm ZAMS}=15$ \msun~(non-rotating) progenitor star, with an oxygen mass of 0.8 \msun. Spectra from 12 and 19 \msun~progenitors under- and overproduce most of the nebular phase lines, respectively, with particular discrepancy for the high-mass model. This result is in fair agreement with the analysis of pre-explosion imaging ($M_{\rm ZAMS}= 11_{-1}^{+2} M_\odot$, Fraser et al. 2012, in prep.), but in disagreement with the progenitor mass inferred from hydrodynamical modeling ($M_{\rm ZAMS}> $25 \msun, \citet{Utrobin2009}).
\item In the $300-600$ day range, we find that the [\ion{O}{i}] 6300, 6364 doublet emits 1, 2, and $4-6$\% of the \iso{56}Co decay energy for 12, 15, and 19 \msun\ progenitor stars. This result may be used for a coarse diagnosis of other Type IIP SNe with similar \iso{56}Ni mass (0.06 \msun).
\item By modeling the mid-infrared iron-group lines observed with Spitzer, we determine a filling factor $f=0.15$ for the Fe/Co/Ni clumps in the ejecta, corresponding to a density $\rho(t) = 7\e{-14}\times (t/100\ \mbox{d})^{-3}$ g cm$^{-3}$. This is similar to the filling factor derived for SN 1987A by \citet{Kozma1998II}, whereas \citet{Li1993iron} likely overestimate the filling factor due to neglect of emission by primordial iron in the hydrogen zone. 
\item By our calculation of the atomic emission lines and continua, we can confirm the contribution by a silicate dust feature to the mid-infrared spectrum, first identified by \citet{Kotak2009}, as well as the presence of SiO and CO fundamental band emission in the spectrum. The CO luminosity matches the total heating of the O/C zone in a $12-15$ \msun~progenitor.

\end{itemize}

\begin{acknowledgements}
We thank the referee for many useful ideas for improving the manuscript. We thank A. Heger for providing the explosion models, and M. Fraser for helpful discussion. Thanks to S. Gelato for advice on code development. This work has been supported by the Swedish Research Council and the Swedish National Space Board. We have made use of the SUSPECT database. 
\end{acknowledgements}

\bibliographystyle{/Users/anders/latex/aa}
\bibliography{/Users/anders/latex/references}

\appendix

\section{Chemical composition of explosion models}

Tables \ref{table:chem12}, \ref{table:chem15}, and \ref{table:chem19} show the chemical composition of the three explosion models we use (taken from \citet{Woosley2007}).

\begin{table*}
\caption{Chemical composition (mass fractions) of the 12 \msun~model used. All values are before any decay of the radioactive isotopes \iso{56}Ni, \iso{57}Co, and \iso{44}Ti has occurred. Mass fractions below $10^{-9}$ are set to zero.}
\label{table:chemcomp1}
\centering

\begin{tabular}{r l l l l l l l l}
  \hline\hline
Element /  Zone  & Fe/He & Si/S & O/Si/S & O/Ne/Mg & O/C & He/C & He/N & H\\
\hline
$^{56}$Ni + $^{56}$Co     & 0.57        & 0.058  & 0           & 0           & 0           & 0           & 0          & 0\\
$^{57}$Co     & 0.034       & $1.5\e{-3}$  & $2.1\e{-6}$ & $2.6\e{-7}$ & $3.8\e{-8}$ & $4.1\e{-9}$ & $1.4\e{-9}$& 0 \\
$^{44}$Ti     & $5.9\e{-4}$ & $1.5\e{-5}$  & 0           & 0           & 0           & 0           & 0          & 0\\
\hline 
H             & $4.0\e{-6}$ & $1.2\e{-6}$ & $2.3\e{-8}$ & $7.0\e{-9}$ & $2.8\e{-9}$ & $2.0\e{-9}$ & $4.7\e{-8}$ & 0.68\\
He            & 0.31        & $8.9\e{-6}$ & $5.3\e{-6}$ & $4.5\e{-6}$ & $2.5\e{-6}$ & 0.73        & 0.99        & 0.31  \\
C             & $3.1\e{-6}$ & $1.6\e{-6}$ & $2.2\e{-4}$ & $3.9\e{-3}$ & 0.19        & 0.23        & $2.1\e{-4}$ & $1.6\e{-3}$ \\
N             & $3.4\e{-6}$ & 0           & $2.9\e{-5}$ & $3.3\e{-5}$ & $2.1\e{-5}$ & $3.1\e{-3}$ & $9.1\e{-3}$ & $2.6\e{-4}$\\
O             & $2.6\e{-5}$ & $1.1\e{-5}$ & 0.80        & 0.66        & 0.74        & 0.020       & $2.2\e{-4}$ & $5.8\e{-3}$\\
Ne            & $2.8\e{-5}$ & $5.0\e{-6}$ & $1.0\e{-4}$ & 0.25        & 0.046       & 0.013       & $1.1\e{-3}$ & $1.3\e{-3}$\\
Na            & $9.9\e{-7}$ & $8.2\e{-7}$ & $1.6\e{-5}$ & $6.2\e{-3}$ & $1.8\e{-4}$ & $2.3\e{-4}$ & $1.6\e{-4}$ & $5.6\e{-5}$\\
Mg            & $5.4\e{-5}$ & $1.2\e{-4}$ & 0.044       & 0.062       & 0.017       & $5.0\e{-3}$ & $7.2\e{-4}$ & $7.3\e{-4}$\\
Al            & $6.1\e{-6}$ & $1.9\e{-4}$ & $3.2\e{-3}$ & $4.0\e{-3}$ & $1.3\e{-3}$ & $5.8\e{-5}$ & $7.3\e{-5}$ & $6.6\e{-5}$\\
Si            & $2.4\e{-4}$ & 0.41        & 0.12        & $4.3\e{-3}$ & $9.2\e{-4}$ & $9.2\e{-4}$ & $8.2\e{-4}$ & $8.2\e{-4}$\\
S             & $2.3\e{-4}$ & 0.39        & 0.020       & $3.0\e{-4}$ & $2.2\e{-4}$ & $3.1\e{-4}$ & $4.2\e{-4}$ & $4.2\e{-4}$\\
Ar            & $2.7\e{-4}$ & 0.054       & $5.7\e{-4}$ & $8.5\e{-5}$ & $8.9\e{-5}$ & $8.2\e{-5}$ & $1.1\e{-4}$ & $1.1\e{-4}$\\
Ca            & $3.1\e{-3}$ & 0.034       & $2.5\e{-5}$ & $3.9\e{-5}$ & $2.5\e{-5}$ & $3.5\e{-5}$ & $7.4\e{-5}$ & $7.4\e{-5}$\\
Sc            & $2.1\e{-7}$ & $2.3\e{-7}$ & $6.7\e{-8}$ & $8.8\e{-7}$ & $9.1\e{-7}$ & $1.0\e{-6}$ & $5.6\e{-8}$ & $4.5\e{-8}$\\
Ti(stable)           & $1.9\e{-3}$ & $5.3\e{-4}$ & $7.5\e{-6}$ & $5.3\e{-6}$ & $8.0\e{-6}$ & $3.2\e{-6}$ & $3.4\e{-6}$ & $3.4\e{-6}$ \\
V             & $6.1\e{-5}$ & $2.2\e{-4}$ & $2.4\e{-6}$ & $3.7\e{-7}$ & $3.7\e{-7}$ & $6.0\e{-7}$ & $4.5\e{-7}$ & $4.3\e{-7}$\\
Cr            & $2.9\e{-3}$ & $6.9\e{-3}$ & $9.1\e{-6}$ & $1.6\e{-5}$ & $1.2\e{-5}$ & $1.5\e{-5}$ & $2.0\e{-5}$ & $2.0\e{-5}$\\
Mn            & $1.9\e{-6}$ & $2.9\e{-4}$ & $1.2\e{-6}$ & $7.6\e{-6}$ & $1.9\e{-6}$ & $3.9\e{-6}$ & $1.6\e{-5}$  & $1.5\e{-5}$\\
Fe (stable)   & $7.0\e{-4}$ & 0.045       & $3.9\e{-4}$ & $9.2\e{-4}$ & $5.1\e{-4}$ & $1.1\e{-3}$ & $1.4\e{-3}$ & $1.4\e{-3}$\\ 
Co (stable)   & $2.2\e{-8}$ & $1.9\e{-9}$ & $1.2\e{-4}$ & $1.7\e{-4}$ & $1.8\e{-4}$ & $2.5\e{-4}$ & $4.3\e{-6}$ & $4.0\e{-6}$\\
Ni(stable)    & 0.029       & $2.5\e{-3}$ & $9.5\e{-4}$ & $3.9\e{-4}$ & $7.2\e{-4}$ & $1.4\e{-4}$ & $8.2\e{-5}$ & $8.2\e{-5}$\\
\hline
 \end{tabular}
\label{table:chem12}
\end{table*}

\begin{table*}
\caption{Same as Table \ref{table:chemcomp1}, for the 15 \msun~model.}
\label{table:chemcomp2}
\centering

\begin{tabular}{r l l l l l l l l}
  \hline\hline
Zone  & Fe/He & Si/S & O/Si/S & O/Ne/Mg & O/C & He/C &  He/N & H\\
 \hline
$^{56}$Ni + $^{56}$Co      & 0.66        & 0.057       & $1.9\e{-7}$ & $3.5\e{-7}$ & $1.1\e{-8}$ & $7.5\e{-8}$ & $2.0\e{-8}$  & 0\\
$^{57}$Co      & 0.033       & $1.4\e{-3}$ & $3.1\e{-6}$ & $2.6\e{-7}$ & $1.3\e{-8}$ & $6.6\e{-9}$ & 0            & 0 \\
$^{44}$Ti      & $5.1\e{-4}$ & $1.5\e{-5}$ & $1.1\e{-9}$ & 0           & 0           & 0           & 0           & 0\\
\hline 
H             & $2.5\e{-6}$ & $4.0\e{-7}$ & $7.4\e{-9}$ & $1.6\e{-9}$ & 0           & 0           & $4.5\e{-8}$ & 0.65\\
He            & 0.23        & $9.5\e{-6}$ & $4.0\e{-6}$ & $3.7\e{-6}$ & $1.6\e{-4}$ & 0.94        & 0.99        & 0.33  \\
C             & $1.4\e{-6}$ & $1.3\e{-6}$ & $3.9\e{-4}$ & $8.5\e{-3}$ & 0.20        & 0.039       & $2.4\e{-4}$ & $1.5\e{-3}$ \\
N             & $2.3\e{-6}$ & 0           & $2.9\e{-5}$ & $6.9\e{-5}$ & $1.3\e{-5}$ & $2.7\e{-3}$ & $9.1\e{-3}$ & $3.0\e{-3}$\\
O             & $1.7\e{-5}$ & $1.0\e{-5}$ & 0.81        & 0.68        & 0.73        & $5.6\e{-3}$ & $1.8\e{-4}$ & $5.4\e{-3}$\\
Ne            & $1.9\e{-5}$ & $7.3\e{-6}$ & $1.4\e{-4}$ & 0.23        & 0.050       & $6.8\e{-3}$ & $1.1\e{-3}$ & $1.2\e{-3}$\\
Na            & $8.6\e{-7}$ & $7.9\e{-7}$ & $2.1\e{-5}$ & $5.2\e{-3}$ & $1.9\e{-4}$ & $1.8\e{-4}$ & $1.8\e{-4}$ & $6.1\e{-5}$\\
Mg            & $4.0\e{-5}$ & $1.4\e{-4}$ & 0.045       & 0.062       & 0.016       & $7.3\e{-4}$ & $7.0\e{-4}$ & $7.3\e{-4}$\\
Al            & $1.0\e{-5}$ & $2.0\e{-4}$ & $4.4\e{-3}$ & $4.0\e{-3}$ & $1.2\e{-4}$ & $7.2\e{-5}$ & $9.0\e{-5}$ & $6.6\e{-5}$\\
Si            & $2.4\e{-4}$ & 0.41        & 0.12        & $4.8\e{-3}$ & $9.4\e{-4}$ & $8.2\e{-4}$ & $8.2\e{-4}$ & $8.2\e{-4}$\\
S             & $2.2\e{-4}$ & 0.39        & 0.019       & $2.9\e{-4}$ & $2.2\e{-4}$ & $4.2\e{-4}$ & $4.2\e{-4}$ & $4.2\e{-4}$\\
Ar            & $2.3\e{-4}$ & 0.055       & $5.9\e{-4}$ & $8.4\e{-5}$ & $8.6\e{-5}$ & $1.1\e{-4}$ & $1.1\e{-4}$ & $1.1\e{-4}$\\
Ca            & $2.7\e{-3}$ & 0.035       & $2.8\e{-5}$ & $3.6\e{-5}$ & $2.6\e{-5}$ & $7.3\e{-5}$ & $7.4\e{-5}$ & $7.4\e{-5}$\\
Sc            & $2.2\e{-7}$ & $2.2\e{-7}$ & $1.2\e{-7}$ & $1.2\e{-6}$ & $1.6\e{-6}$ & $7.3\e{-8}$ & $4.5\e{-8}$ & $4.5\e{-8}$\\
Ti (stable)   & $1.6\e{-3}$ & $5.6\e{-4}$ & $8.1\e{-6}$ & $5.6\e{-6}$ & $7.0\e{-6}$ & $3.4\e{-6}$ & $3.4\e{-6}$ & $3.4\e{-6}$ \\
V             & $3.3\e{-5}$ & $1.5\e{-4}$ & $3.0\e{-6}$ & $4.1\e{-7}$ & $4.5\e{-7}$ & $4.7\e{-7}$ & $4.3\e{-7}$ & $4.3\e{-7}$\\
Cr            & $2.3\e{-3}$ & $6.6\e{-3}$ & $8.2\e{-6}$ & $1.5\e{-5}$ & $1.2\e{-5}$ & $2.0\e{-5}$ & $2.0\e{-5}$ & $2.0\e{-5}$\\
Mn            & $1.7\e{-6}$ & $2.8\e{-4}$ & $1.7\e{-6}$ & $6.7\e{-6}$ & $2.2\e{-6}$ & $1.7\e{-5}$ & $1.5\e{-5}$ & $1.5\e{-5}$\\
Fe (stable)   & $7.6\e{-4}$ & 0.042       & $3.4\e{-4}$ & $8.5\e{-4}$ & $5.5\e{-4}$ & $1.4\e{-3}$ & $1.4\e{-3}$ & $1.4\e{-3}$\\
Co (stable)   & $2.3\e{-8}$ & $2.0\e{-9}$ & $1.9\e{-4}$ & $1.8\e{-4}$ & $2.0\e{-4}$ & $4.8\e{-6}$ & $4.0\e{-6}$ & $4.0\e{-6}$\\
Ni (stable)   & 0.028       & $2.3\e{-3}$ & $8.6\e{-4}$ & $4.4\e{-4}$ & $6.7\e{-4}$ & $8.2\e{-5}$ & $8.2\e{-5}$ & $8.2\e{-5}$\\
\hline
 \end{tabular}
\label{table:chem15}
\end{table*}

\begin{table*}
\caption{Same as Table \ref{table:chemcomp1}, for the 19 \msun~model.}
\label{table:chemcomp3}
\centering

\begin{tabular}{r l l l l l l l l}
  \hline\hline
Zone           & Fe/He       & Si/S       & O/Si/S       & O/Ne/Mg     & O/C         & He/C       & He/N & H\\
 \hline
$^{56}$Ni+$^{56}$Co & 0.76    & 0.029       & $5.8\e{-7}$ & $1.3\e{-7}$ & $9.4\e{-9}$ & $5.0\e{-8}$ & $1.6\e{-8}$ & 0\\
$^{57}$Co      & 0.033       & $1.0\e{-3}$ & $1.1\e{-5}$ & $9.8\e{-8}$ & $1.9\e{-8}$ & $3.7\e{-9}$ & 0           & 0 \\
$^{44}$Ti      & $2.5\e{-4}$ & $1.1\e{-5}$ & $4.8\e{-6}$ & 0           & 0           & 0           & 0           & 0\\
\hline 
H             & $2.5\e{-6}$ & $4.1\e{-8}$ & $8.6\e{-9}$ & 0           & 0           & 0           & $3.7\e{-8}$ & 0.62\\
He            & 0.13        & $8.5\e{-6}$ & $4.3\e{-6}$ & $2.0\e{-6}$ & 0.078       & 0.96        & 0.99        & 0.37  \\
C             & $3.2\e{-7}$ & $1.3\e{-6}$ & $8.5\e{-5}$ & 0.031       & 0.28        & 0.023       & $2.6\e{-4}$ & $1.4\e{-3}$ \\
N             & $1.4\e{-6}$ & 0           & $1.8\e{-5}$ & $3.2\e{-5}$ & $4.8\e{-6}$ & $6.4\e{-4}$ & $9.0\e{-3}$ & $3.7\e{-3}$\\
O             & $9.2\e{-6}$ & $1.3\e{-5}$ & 0.45        & 0.70        & 0.62        & $3.3\e{-3}$ & $1.7\e{-4}$ & $4.9\e{-3}$\\
Ne            & $9.6\e{-6}$ & $1.6\e{-6}$ & $1.4\e{-4}$ & 0.22        & 0.017       & 0.011       & $1.1\e{-3}$ & $1.2\e{-3}$\\
Na            & $6.2\e{-7}$ & $1.2\e{-6}$ & $1.2\e{-6}$ & $7.0\e{-3}$ & $1.9\e{-4}$ & $1.8\e{-4}$ & $1.8\e{-4}$ & $7.2\e{-5}$\\
Mg            & $8.9\e{-6}$ & $1.7\e{-4}$ & $7.2\e{-4}$ & 0.042       & $4.9\e{-3}$ & $7.2\e{-4}$ & $7.0\e{-4}$ & $7.3\e{-4}$\\
Al            & $2.1\e{-5}$ & $2.6\e{-4}$ & $1.9\e{-4}$ & $4.0\e{-3}$ & $6.7\e{-5}$ & $7.5\e{-5}$ & $9.8\e{-5}$ & $6.7\e{-5}$\\
Si            & $1.5\e{-4}$ & 0.45        & 0.27        & $3.1\e{-3}$ & $8.9\e{-4}$ & $8.2\e{-4}$ & $8.2\e{-4}$ & $8.2\e{-4}$\\
S             & $1.1\e{-4}$ & 0.39        & 0.21        & $2.6\e{-4}$ & $3.2\e{-4}$ & $4.2\e{-4}$ & $4.2\e{-4}$ & $4.2\e{-4}$\\
Ar            & $1.0\e{-4}$ & 0.049       & 0.043       & $8.4\e{-5}$ & $8.9\e{-5}$ & $1.1\e{-4}$ & $1.1\e{-4}$ & $1.1\e{-4}$\\
Ca            & $1.3\e{-3}$ & 0.029       & 0.013       & $3.1\e{-5}$ & $5.0\e{-5}$ & $7.3\e{-5}$ & $7.4\e{-5}$ & $7.4\e{-5}$\\
Sc            & $2.6\e{-7}$ & $2.1\e{-7}$ & $1.2\e{-6}$ & $1.6\e{-6}$ & $5.5\e{-7}$ & $7.6\e{-8}$ & $4.5\e{-8}$ & $4.5\e{-8}$\\
Ti (stable)   & $8.1\e{-4}$ & $3.7\e{-4}$ & $9.3\e{-5}$ & $6.2\e{-6}$ & $4.4\e{-6}$ & $3.4\e{-6}$ & $3.4\e{-6}$ & $3.4\e{-6}$ \\
V             & $2.1\e{-5}$ & $1.5\e{-4}$ & $3.7\e{-6}$ & $4.1\e{-7}$ & $3.4\e{-7}$ & $4.8\e{-7}$ & $4.3\e{-7}$ & $4.3\e{-7}$\\
Cr            & $1.3\e{-3}$ & $4.5\e{-3}$ & $4.5\e{-5}$ & $1.3\e{-5}$ & $1.7\e{-5}$ & $2.0\e{-5}$ & $2.0\e{-5}$ & $2.0\e{-5}$\\
Mn            & $2.7\e{-6}$ & $2.7\e{-4}$ & $2.2\e{-6}$ & $3.8\e{-6}$ & $9.4\e{-6}$ & $1.7\e{-5}$ & $1.5\e{-5}$ & $1.5\e{-5}$\\
Fe (stable)   & $1.1\e{-3}$ & 0.050       & $4.5\e{-4}$ & $6.9\e{-4}$ & $1.1\e{-3}$ & $1.4\e{-3}$ & $1.4\e{-3}$ & $1.4\e{-3}$\\
Co (stable)   & $3.6\e{-8}$ & $4.4\e{-9}$ & $1.4\e{-6}$ & $1.9\e{-4}$ & $1.3\e{-4}$ & $4.4\e{-6}$ & $4.0\e{-6}$ & $4.0\e{-6}$\\
Ni (stable)   & 0.038       & $2.4\e{-3}$ & $1.1\e{-3}$ & $5.6\e{-4}$ & $2.3\e{-4}$ & $8.2\e{-5}$ & $8.2\e{-5}$ & $8.2\e{-5}$\\
\hline
 \end{tabular}
\label{table:chem19}
\end{table*}

\section{Code updates compared to J11}
\label{sec:updates}


\subsection{Photoexcitation/deexcitation rates}
Because the computations here are for earlier epochs than in J11, photoexcitation/deexcitation rates in the NLTE solutions have to be included. We therefore implemented a modified Monte Carlo operator to allow for partial deposition of photoexcitation energy along the photon trajectories. A photon packet of comoving energy $E$ and frequency $\nu$, coming into resonance with a transition $l\rightarrow u$ of atom $k$, where the upper level $u \le u_{\rm max}^{\rm k}$ (=the number of levels included in the NLTE solution of atom $k$), now increases the number of photoabsorbed packages by
\begin{equation}
\Delta N_{\rm i,k,l,u}= \frac{E}{h\nu}(1-e^{-\tau_{\rm lu}^{\rm S}})~,
\label{eq:pa}
\end{equation}
where $i$ is the zone index and $\tau_{\rm lu}^{\rm S}$ the Sobolev optical depth. 
The packet then continues (undeflected) with energy $E e^{-\tau_{\rm lu}^{\rm S}}$. 
For lines with $u>u_{\rm max}^{\rm k}$, we retain the treatment described in J11 (scattering/fluorescence is treated explicitly in the transfer). 

The values for $u_{\rm max}^{\rm k}$ should be chosen large enough to reasonably well sample the total number of photoexcitations from each level, but small enough to ensure stability and keep computation times limited (matrix inversion times in the statistical equilibrium solutions scale as $u_{\rm max}^{\rm k}$ cubed). The $u_{\rm max}^{\rm k}$ values used here can be found in Table \ref{table:modelatoms}. We have chosen to set $u_{\rm max}^{\rm k}=300$ for all iron-group elements (except \ion{Fe}{i}, \ion{Fe}{ii} and \ion{Cr}{ii}, where we solve for more levels), and $u_{\rm max}^{\rm k}=200$ for the non-iron-group elements.

In the Sobolev approximation, the photoexcitation rates to be used in the statistical equilibrium equations are given by \citep{Castor1970}
\begin{equation}
R_{\rm lu} = B_{\rm lu}\beta_{\rm lu} J_{\rm \nu}^{\rm b}~,
\label{eq:photoexc}
\end{equation}
where $B_{\rm lu}$ is the Einstein coefficient for photoexcitation, $\beta_{\rm lu}$ is the escape probability, and $J_{\rm \nu}^{\rm b}$ is the far blue-wing mean intensity (before $J_\nu$ is affected by the line). The deexcitation rate is related by detailed balance
\begin{equation}
R_{\rm ul} = \frac{g_{\rm l}}{g_{\rm u}}R_{\rm lu}~,
\label{Eq:einsteinrel}
\end{equation}
where $g_{\rm l}$ and $g_{\rm u}$ are the statistical weights.

Overlapping lines may cause $J_\nu^{\rm b}$ to significantly vary with the choice of the blue frequency. We therefore do not compute the photoexcitation rates from this equation, but by adding the contributions from Eq. (\ref{eq:pa}). Since stimulated emission is treated as negative absorption in the Sobolev optical depth, Eq. (\ref{eq:pa}) gives only the \emph{net} number of photons absorbed in the line; absorptions minus stimulated emissions. The actual number of absorptions occurring is therefore higher by a factor $(1-g_{\rm l} n_{\rm u}/g_{\rm u} n_{\rm l})^{-1}$, where $n_{\rm l}$ and $n_{\rm u}$ are the populations of levels $l$ and $u$, so the estimator for the photoexcitation rate is 
\begin{equation}
R_{\rm i,k,l,u} = \frac{N_{\rm i,k,l,u}}{n_{\rm i,k,l}}\left(1 - \frac{g_{\rm k,l} n_{\rm i,k,u}}{g_{\rm k,u} n_{\rm i,k,l}}\right)^{-1}~,
\end{equation}
where $N_{\rm i,k,l,u}$ is the total number of packets absorbed in the transition in the Monte Carlo simulation. 

For lines with $\tau_{\rm lu}^{\rm s} <0$, we allow lasering to occur in the transfer by using the same formalism as for normal positive optical depths (Eq. (\ref{eq:pa})). In these cases we compute the \emph{deexcitation} rate as
\begin{equation}
R_{\rm i,k,u,l} = \frac{|N_{\rm i,k,l,u}|}{n_{\rm i,k,u}}\left(1 - \frac{g_{\rm k,u} n_{\rm i,k,l}}{g_{\rm k,l} n_{\rm i,k,u}}\right)^{-1}~,
\end{equation}
and then use Eq. (\ref{Eq:einsteinrel}) to get the excitation rate.

\subsection{Charge transfer}
We now generalize the treatment of the \ion{O}{ii} + \ion{C}{i} $\rightarrow$\ion{O}{i}($2p(^1D$)) + \ion{C}{ii} reaction in J11, by assuming that all charge transfer reactions occur to the state (either in the recombining or in the ionizing element) that gives the smallest energy defect.

\subsection{Photoionization rates and photoelectric heating rates}
We have also modified the estimators for the photoionization rates and the photoelectric heating rates. 
Whereas the previous treatment did not allow splitting of packets upon continuum interaction, a photon packet passing a region of optical depth $\tau$, of which photoionization by atom $k$, level $l$, contributes $\tau_{\rm pi}^{\rm k,l}$, now adds a contribution
\begin{equation}
\Delta N_{\rm i,k,l} = \frac{\tau_{\rm pi}^{\rm k,l}}{\tau}\frac{E}{h\nu}\left(1-e^{-\tau}\right)~,
\end{equation}
to the number of photons $N_{\rm i,k,l}$ absorbed in the corresponding continuum. The heating rate is increased by
\begin{equation}
\Delta H_{\rm pi}^{\rm i} = \frac{1}{V_{\rm i}} \sum_{\rm k,l} \Delta N_{\rm i,k,l}\left(h\nu - \chi_{\rm k,l}\right)~,
\end{equation}
where $V_{\rm i}$ is the volume of the zone, and $\chi_{\rm k,l}$ is the ionization potential.

\subsection{Electron scattering}
We implement electron scattering in the following way. The total continuum optical depth to the next point of interest (a zone boundary, a line, or a photoionization threshold) is
\begin{equation}
\Delta \tau_{\rm cont} = \Delta \tau_{\rm pi} + \Delta \tau_{\rm d} + \Delta \tau_{\rm T}~,
\end{equation}
where $\Delta \tau_{\rm pi}$ is the photoionization optical depth, $\Delta \tau_{\rm d}$ is the dust optical depth, and $\Delta \tau_{\rm T}$ is the Thomson optical depth.
The fraction of photons absorbed in photoionization continua is
\begin{equation}
f_{\rm pi} = \frac{\Delta \tau_{\rm pi}}{\Delta \tau_{\rm cont}}\left(1-e^{-\Delta \tau_{\rm cont}}\right)~,
\end{equation}
the fraction absorbed by dust is
\begin{equation}
f_{\rm d} = \frac{\Delta \tau_{\rm d}}{\Delta \tau_{\rm cont}}\left(1-e^{-\Delta \tau_{\rm cont}}\right)~,
\end{equation}
the fraction that Thomson scatter is
\begin{equation}
f_{\rm T} = \frac{\Delta \tau_{\rm T}}{\Delta \tau_{\rm cont}}\left(1-e^{-\Delta \tau_{\rm cont}}\right)~,
\end{equation}
and the fraction that do not interact is
\begin{equation}
f_{\rm pass} = e^{-\Delta \tau_{\rm cont}} = 1 - f_{\rm pi} - f_{\rm d} - f_{\rm T}~.
\end{equation}
We let the fractions $f_{\rm pi}$ and $f_{\rm d}$ be absorbed in photoionization continua and by dust, respectively. We then let the remaining fraction ($f_{\rm T}+f_{\rm pass}$) Thomson scatter with probability 
\begin{equation}
p_{\rm T} = \frac{f_{\rm T}}{f_{\rm T} + f_{\rm pass}}~,
\end{equation}
and continue through with probability $1-p_{\rm T}$. If Thomson scattering occurs, we draw a position for the scattering from
\begin{equation}
\Delta \tau = -\ln\left[1 + z\left(e^{-\Delta \tau_{\rm cont}}-1\right)\right]~,
\end{equation}
and a new direction cosine $\mu$ from
\begin{equation}
z = \frac{\int_{-1}^\mu p(\mu') d\mu'}{\int_{-1}^1 p(\mu') d\mu'}~,
\end{equation}
where $z$ is a uniform random number between 0 and 1, and $p(\mu')$ is the Rayleigh phase function
\begin{equation}
p(\mu') = \frac{3}{4}\left(1+\mu'^2\right)~.
\end{equation}

We treat the scattering as coherent in the fluid frame, i.e. we ignore the small energy shifts occurring due to the thermal motions of the electrons, as well as the Compton shift.

\subsection{Line overlap}
\label{sec:lineoverlap}
When the density is high, the line escape probabilities may be enhanced from their Sobolev values $\beta^{\rm S}$ by overlapping continua and lines, so that
\begin{equation}
\beta^{\rm eff} \simeq \beta^{\rm S} + \beta^{\rm C} + \beta^{\rm L}~.
\end{equation}
We take the $\beta^{\rm C}$ term into account as described in J11. In this paper, we also take the $\beta^{\rm L}$ term into account for Ly$\alpha$ and Ly$\beta$, which are important for the spectral formation in Type II SNe. 

Ly$\alpha$ reaches optical depths as high as $\tau^{\rm S} = 10^{10}-10^{11}$ in the early nebular phase, and may branch into several iron-group lines around 1215.67 \AA~before escaping the resonance layer. Preliminary simulations (which we hope to expand on in a future publication) show that at typical conditions in the early nebular phase, the branching probability is
\begin{equation}
\beta_{\rm Ly\alpha}^{\rm L} \approx 10^{-9}-10^{-8}~.
\end{equation}
Here we use $\beta_{\rm Ly\alpha}^{\rm L} = 5\e{-9}$. We assume that the absorbing lines are the ones which, at 300 days in the core H region, are optically thick and within 100 Doppler widths (on the red side) of the Ly$\alpha$ rest wavelength, with allocation to these transitions in equal proportions. The lines, which fall in the $1215.67-1220$ \AA~range, are listed in Table \ref{table:lyaalloc}.

\begin{table}[htb]
\centering
\caption{Transitions into which (part of the) Ly$\alpha$ flux is allocated.}
\begin{tabular}{c|c}
\hline
\hline
\ion{Fe}{ii} transitions:                              &  1218.70 $4s(a^4D_{1/2}) - 5p(^4P ^o_{5/2})$\\
 1217.85 $4s(a^4D_{7/2}) - (5D)5p_{9/2}$ &  1219.41  $3d(7a^2G_{9/2}) - 4p(^2H^o_{9/2})$\\
 1217.15 $4s(a^4D_{3/2}) - 4p(^4P^o_{1/2})$  &  1219.80  $4s(a^4H_{13/2}) -\ ^3P^o(^4I^o_{15/2})$\\
 1215.98 $4s(a^4D_{5/2}) - 4p(^4S^o_{3/2})$  &   \ion{Cr}{ii} transitions:\\
 1215.85  $4s(a^4D_{5/2}) - 5p(^4D^o_{5/2})$ &  1216.31 $4s(a^6D_{5/2}) - 5p(^6P_{7/2})$\\
 1216.24 $4s(a^4D_{3/2}) - 5p(^4P^o_{5/2})$  &  1217.31 $4s(a^6D_{7/2}) - sp(x^6D_{7/2})$\\
 1216.27   $4s(a^4D_{3/2}) - 4p(^4P ^o_{3/2})$& 1218.63 $4s(a^6D_{7/2}) - 5p(^6P_{7/2})$ \\
 1216.52 $4s(a^4D_{1/2}) - 5p(^4D ^o_{3/2})$  & 1217.14  $4s(a^6D_{9/2}) - sp(x^6D_{9/2})$ \\
 1216.85  $4s(b^4F_{9/2}) -\ ^3P ^o(^4H ^o_{11/2})$ &  1217.85 $4s(a^6D_{3/2}) - 5p(^6P_{5/2})$\\
 1217.62 $4s(b^4F_{7/2}) -\ ^3P^o(^4H ^o_{9/2})$  &  1218.91 $4s(a^6D_{1/2}) - 5p(^6P_{3/2})$ \\
 1218.09 $4s(a^4D_{3/2}) - 5p(^6F^o_{5/2})$ &  1219.56 $4s(a^6D_{5/2}) - 5p(^6P_{5/2})$\\
 1218.19 $4s(b^4F_{5/2}) -\ ^3P^o(^4H ^o_{7/2})$&  1219.96 $4s(a^6D_{3/2}) - 5p(^6P_{3/2})$ \\
\hline
\end{tabular}
\label{table:lyaalloc}
\end{table}

 The \ion{O}{i} \wl1025.76 $2p^4(^3P) - 3d(^3D^o)$  transition lies within the Doppler core of Ly$\beta$. If the \ion{O}{i} \wl1.129 $\mu$m $3p(^3P)-3d(^3D^o)$ transition is optically thin, and complete $l$-mixing in both \ion{H}{i} $n=3$ and \ion{O}{i} $3d^3D$ is assumed, the branching probability for \ion{O}{i} fluorescence is\footnote{5/9 of \ion{O}{i} can absorb the line, there is a $\sim$30\% chance that fluorescence occurs in \ion{O}{i}, the $Ag_{\rm u}/g_{\rm l}$ values are $1.3\e{8}$ for the \ion{O}{i} line and $5.0\e{9}$ for Ly$\beta$ $\rightarrow P = 5/9\cdot 0.3\cdot 1.3\e{8}/5.0\e{8}=0.043$.}
\begin{equation}
\beta_{\rm Ly\beta}^{\rm L} = 0.043\frac{\mbox{n(\ion{O}{i})}}{\mbox{n(\ion{H}{i})}}~,
\end{equation}
where n(\ion{O}{i}) and n(\ion{H}{i}) are the number densities of \ion{O}{i} and \ion{H}{i}. At solar metallicity $\mbox{n(O)}/\mbox{n(H)}=4.9\e{-4}$, so $\beta_{\rm Ly\beta}^{\rm L}$ = $2.1\e{-5}$, which is the value we use.

\subsection{Photoionization cross sections}
For excited states with unknown photoionization cross sections, we use
the hydrogenic approximation \citep{RL}
\begin{equation}
\sigma_{\rm k,l} = 7.9\e{-18}\left(\frac{\nu}{\nu^{\rm 0}_{\rm k,l}}\right)^{-3}g_{\rm k,l} n_{\rm k,l} Z_{\rm k}^{-2}~\mbox{cm}^2~,
\end{equation}
where $\nu^{0}_{\rm k,l} $ is the ionization threshold, $g_{\rm k,l}$ is the Gaunt factor (which we set to unity for all levels), $Z_{\rm k}$ is the effective nuclear charge (number of protons minus number of screening electrons), and $n_{\rm k,l}$ is the effective principal quantum number
\begin{equation}
n_{\rm k,l} = \left(1-\frac{E_{\rm k,l}}{\chi_{\rm k,0}}\right)^{-1/2}~,
\end{equation}
where $E_{\rm k,l}$ is the level energy and $\chi_{\rm k,0}$ is the ground state ionization potential.

We use a threshold for how many levels to compute photoionization cross sections for. For this paper this limit was set to 50 levels. Tests showed that contributions by levels above this were small for all epochs modeled.

\subsection{Free-bound emissivities}
For hydrogen, we use detailed emissivity functions obtained by applying the Milne relations to the photoionization cross sections, which we obtain from \citet{Verner1996} for $n=1$, and \citet{Bethe1957} for $n=2$. For $n=3$ and higher we use the hydrogenic approximation.

For all other atoms $k$ we use a simplified prescription, where recombination to level $l$ gives an emissivity of 
\begin{equation}
4\pi j_{\rm k,l}^{\rm f-b} = n_{\rm e} n_{\rm +} \alpha_{\rm k,l}(T)\left(\xi_{\rm k,l}(T)\frac{3}{2}kT + \chi_{\rm k,l}\right)~,
\end{equation}
is emitted over a box-shaped profile between $0.5-1.5$ $kT$. Here $\alpha_{\rm k,l}(T)$ is the specific recombination coefficient, and $\xi_{\rm k,l}(T)$ is taken as 0.55 for all atoms, levels,  and temperatures, the value for hydrogen at a few thousand K \citep{Spitzer1948}.

\subsection{Collision strengths}
For transitions with unknown collision strengths we use van Regemorter's formula \citep{Regemorter1962} for allowed lines (taken as $f_{\rm osc} > 10^{-3}$), and
\begin{equation}
\Omega_{\rm lu} = 0.004 g_{\rm l} g_{\rm u}~,
\end{equation}
for forbidden lines \citep{Axelrod1980}. 

\subsection{Model atoms}
Table \ref{table:modelatoms} lists the model atoms used. Compared to J11, we now include more levels in the NLTE solutions, and have also added \ion{Fe}{iii} and \ion{Co}{iii}\footnote{Data from \texttt{http://kurucz.harvard.edu/atoms.html}}. We have added collision strengths for \ion{Ni}{ii} \citep [][we use the values at 5000 K]{Cassidy2010, Cassidy2011}. The sources for the rest of the atomic data are given in J11.

\begin{table}[htbp]
\caption{Model atoms used. The \emph{Levels} column shows the number
  of levels (including spin-orbit splitting) included in the NLTE solutions ($u_{\rm max}$), followed by the number included in the radiative transfer. The column to the right identifies the corresponding levels. See J11 for references for the atomic data.}
\centering
\begin{tabular}{c c c}
\hline
\hline
Element & Levels & Highest term\\
\hline
\ion{H}{i} & 210 / 465 & n=20 / n=30\\
\ion{He}{i} & 29 / 29 & $5p(^1P ^o)$\\
\ion{C}{i} & 90 / 90 & $5d(^1D^o$)\\
\ion{C}{ii} & 77 / 77 & $8g(^2G)$\\
\ion{N}{i} & 68 / 68 & $5s(^2P)$\\
\hline
\ion{N}{ii} & 148 / 148 & $3p(D^o)$\\
\ion{O}{i} & 135 / 135 & $8d(^3D^o)$ \\
\ion{O}{ii} & 16 / 16 & $2s.2p^4(^2S)$\\
\ion{Ne}{i} & 3 / 3 & $3s(^2[3/2]^o)$ \\
\ion{Ne}{ii} & 3 / 3 & $2p^6(^2S)$\\
\hline
\ion{Na}{i} & 66 / 66 & $9f(^2F^o)$ \\
\ion{Mg}{i} & 200 / 371 & $37d(^3D) / 80d(^3D)$\\
\ion{Mg}{ii} & 35 / 35 & $7p(^2P ^o)$\\
\ion{Al}{i} & 86 / 86 & $21s(^2S)$\\
\ion{Al}{ii} & 154 / 154 & $3p3d(^3D^o)$\\
\hline
\ion{Si}{i} & 200 / 494 & $10s(^3P) / 21s(^1P)$\\
\ion{Si}{ii} & 77 / 77 & $9p(^2P^o)$\\
\ion{S}{i} & 125 / 125 & $8f(^3F)$\\
\ion{S}{ii} & 5 / 5 & $3p^3(^2P\rm ^o)$\\
\ion{Ar}{i} & 3 / 3 & $4s(^2[3/2]^o)$\\
\hline
\ion{Ar}{ii} & 2 / 2 & $3p^5(^2P ^o)$\\
\ion{Ca}{i} & 198 / 198 & $16d(^3D)$\\
\ion{Ca}{ii} & 69 / 69 & $16d(^2D)$\\
\ion{Sc}{i} & 254 / 254 & $7p(^2P ^o)$\\
\ion{Sc}{ii} & 165 / 165 & $6f(^1H ^o)$\\
\hline
\ion{Ti}{i} & 300 / 394 & $5p(^5F) / 4p^2(e^1P)$\\
\ion{Ti}{ii} & 212 / 212 & $5d(^4H)$\\
\ion{V}{i} & 300 / 502 & $4p(t^4G ^o) / 4p^2(r^2H^o)$\\
\ion{V}{ii} & 300 / 323 & $F_{7/2}4f(^2[11/2] ^o) / F_{9/2}4f(^2[11/2] ^o)$\\
\ion{Cr}{i} & 300 / 392 & $4p(t^5P ^o) / 21p(e^5F)$ \\
\hline
\ion{Cr}{ii} & 364\tablefootmark{a} / 500 & $4s4p(x^6D ^o) / 4f(^4H^o)$\\
\ion{Mn}{i} & 293 / 293 & $4p(z^2I ^o)$\\
\ion{Mn}{ii} & 300 / 509 & $4s4p(w^5D^o) / 9h(^7H^o)$\\
\ion{Fe}{i} & 496 / 496 & $4s4p(t^3H ^o)$\\
\ion{Fe}{ii} & 533\tablefootmark{b} / 578 & $4s4p(^4H ^o) / 4d(^8D)$\\
\hline
\ion{Fe}{iii} & 300 / 600 & $^4P(^1F ^o) / 5p(^5F ^o)$\\
\ion{Co}{i} & 300 / 317 & $4d(e^6H) / 5s(37^o)$\\
\ion{Co}{ii} & 108 / 108 & $7d(^5H)$\\
\ion{Co}{iii} & 300 / 306 & $4d(e^4F) / 5s(f^4D)$\\  
\ion{Ni}{i} & 136 / 136 & $4d(^2[1/2])$ \\
\ion{Ni}{ii} & 300 / 500 & $5p(^4S^o) / 6g(^2[8])$ \\
\hline
\end{tabular}
\tablefoot{
  \tablefoottext{a}{Ly$\alpha$ pumping occurs up to level 364.}
\tablefoottext{b}{ Ly$\alpha$ pumping occurs up to level 533.}
}
\label{table:modelatoms}
\end{table}

\end{document}